\documentclass[manuscript, acmsmall, screen, nonacm]{acmart}

\AtBeginDocument{%
  \providecommand\BibTeX{{%
    Bib\TeX}}}

\setcopyright{acmcopyright}
\copyrightyear{2024}
\acmYear{2024}
\acmDOI{XXXXXXX.XXXXXXX}

\acmJournal{TOSEM}

\usepackage{indentfirst} 
\usepackage{algorithmic}
\usepackage{graphicx}
\usepackage{textcomp}
\usepackage{xcolor}
\usepackage{enumerate}
\usepackage{listings}
\usepackage{color}
\usepackage{xcolor}
\usepackage{booktabs}
\usepackage{xspace}
\usepackage{subcaption}
\usepackage{verbatim}
\usepackage{mathtools}
\usepackage{tcolorbox}
\definecolor{dkgreen1}{rgb}{0,0.6,0}
\definecolor{gray1}{rgb}{0.5,0.5,0.5}
\definecolor{mauve1}{rgb}{0.58,0,0.82}
\usepackage{url}

\newcommand{\tool}{{{PopSweeper}}\xspace}
\newcommand{\toolClassify}{{{PopSweeper$_{\text{classify}}$}}\xspace}
\newcommand{\toolDetect}{{{PopSweeper$_{\text{detect}}$}}\xspace}

\newcommand{\rqboxc}[1]{\begin{tcolorbox}[left=4pt,right=4pt,top=4pt,bottom=4pt,colback=gray!35,colframe=gray!35,before skip=3pt,after skip=3pt]#1\end{tcolorbox}}

\newcommand{\peter}[1]{\textcolor{red}{{\it [Peter says: #1]}}}
\newcommand{\liu}[1]{\textcolor{blue}{{\it [wei says: #1]}}}

\newcommand{\phead}[1]{\vspace{0.8mm} \noindent {\bf #1}}

\def\BibTeX{{\rm B\kern-.05em{\sc i\kern-.025em b}\kern-.08em
    T\kern-.1667em\lower.7ex\hbox{E}\kern-.125emX}}
\begin{document}

\title{\tool: Automatically Detecting and Resolving App-Blocking Pop-Ups to Assist Automated Mobile GUI Testing}

\author{Linqiang Guo}
\affiliation{%
  \institution{Software PErformance, Analysis, and Reliability (SPEAR) lab, Concordia University}
  \city{Montreal}
  \state{Quebec}
  \country{Canada}}
\email{g_linqia@live.concordia.ca}
\orcid{0009-0004-7456-3145}

\author{Wei~Liu}
\affiliation{%
  \institution{Software PErformance, Analysis, and Reliability (SPEAR) lab, Concordia University}
  \city{Montreal}
  \state{Quebec}
  \country{Canada}}
\email{w\_liu201@encs.concordia.ca}
\orcid{0000-0001-8956-730X}

\author{Yi Wen Heng}
\affiliation{%
  \institution{Software PErformance, Analysis, and Reliability (SPEAR) lab, Concordia University}
  \city{Montreal}
  \state{Quebec}
  \country{Canada}}
\email{he\_yiwen@encs.concordia.ca}
\orcid{0009-0003-7371-3319}

\author{Tse-Hsun~(Peter)~Chen}
\affiliation{%
  \institution{Software PErformance, Analysis, and Reliability (SPEAR) lab, Concordia University}
  \city{Montreal}
  \state{Quebec}
  \country{Canada}}
\email{peterc@encs.concordia.ca}
\orcid{0000-0003-4027-0905}

\author{Yang Wang}
\affiliation{%
  \institution{Concordia University}
  \city{Montreal}
  \state{Quebec}
  \country{Canada}}
\email{yang.wang@concordia.ca}

\renewcommand{\shortauthors}{Guo et al.}
\begin{abstract}
Graphical User Interfaces (GUIs) are the primary means by which users interact with mobile applications, making them crucial to both app functionality and user experience. 
However, a major challenge in automated testing is the frequent appearance of app-blocking pop-ups, such as ads or system alerts, which obscure critical UI elements and disrupt test execution, often requiring manual intervention. These interruptions lead to inaccurate test results, increased testing time, and reduced reliability, particularly for stakeholders conducting large-scale app testing. To address this issue, we introduce \tool, a novel tool designed to detect and resolve app-blocking pop-ups in real-time during automated GUI testing. \tool combines deep learning-based computer vision techniques for pop-up detection and close button localization, allowing it to autonomously identify pop-ups and ensure uninterrupted testing. We evaluated \tool on over 72K app screenshots from the RICO dataset and 87 top-ranked mobile apps collected from app stores, manually identifying 832 app-blocking pop-ups. \tool achieved 91.7\% precision and 93.5\% recall in pop-up classification and 93.9\% BoxAP with 89.2\% recall in close button detection. Furthermore, end-to-end evaluations demonstrated that PopSweeper successfully resolved blockages in 87.1\% of apps with minimal overhead, achieving classification and close button detection within 60 milliseconds per frame. These results highlight \tool’s capability to enhance the accuracy and efficiency of automated GUI testing by mitigating pop-up interruptions.

\end{abstract}

\ccsdesc[500]{Software and its engineering~Software testing and debugging}
\keywords{Mobile App, In-app Ads, Mobile Pop-ups, Automated GUI Testing.}

\maketitle

\section{Introduction}
\label{sec:intro}

Graphical User Interfaces (GUIs) are the sole means users interact with mobile applications (i.e., apps), making them central to the app's functionality and user experience. 
Hence, ensuring reliable operations when users interact with apps is critical to delivering a high-quality user experience. 
To ensure app quality, automated GUI testing has become an integral part of app development~\cite{2023_ICSE_Efficiency_Matters_Speeding_Up_Automated_Testing_with_GUI_Rendering_Inference}. Testers write test scripts that simulate user interactions and systematically test app functionality across a variety of devices and environments~\cite{KroppAutomated_GUI_Testing2010}. 
By automating the process of tapping, swiping, and navigating through the app's interface, developers can quickly detect bugs, performance issues, and crashes that might affect user experience.

Automated GUI testing relies heavily on accurately identifying and interacting with UI elements, as the test scripts need to know the precise location of buttons, text fields, and other interface components to simulate user actions correctly~\cite{2020_FSE_Object_detection_GUI, 2021_FSE_GUI_widget_detection_for_mobile_games}. One common method for automating this process is through record-and-replay tools, which capture and replay user interactions, or coordinate-based testing, where scripts interact with elements using specific screen coordinates.
Hence, accurate localization is crucial for ensuring test scripts execute as expected. Any changes in the location or even shape of the UI elements can deviate the tests, leading to inaccurate results~\cite{2019_EMSE_Scripted_GUI_testing_of_Android_open_source_apps, 2022_TSE_GUI_Guided_Test_Script_Repair, 2024_ICSE_Comprehensive_Semantic_Repair_of_Obsolete_GUI_Test_Scripts}. 

\begin{figure}
	\includegraphics[width=1\textwidth]{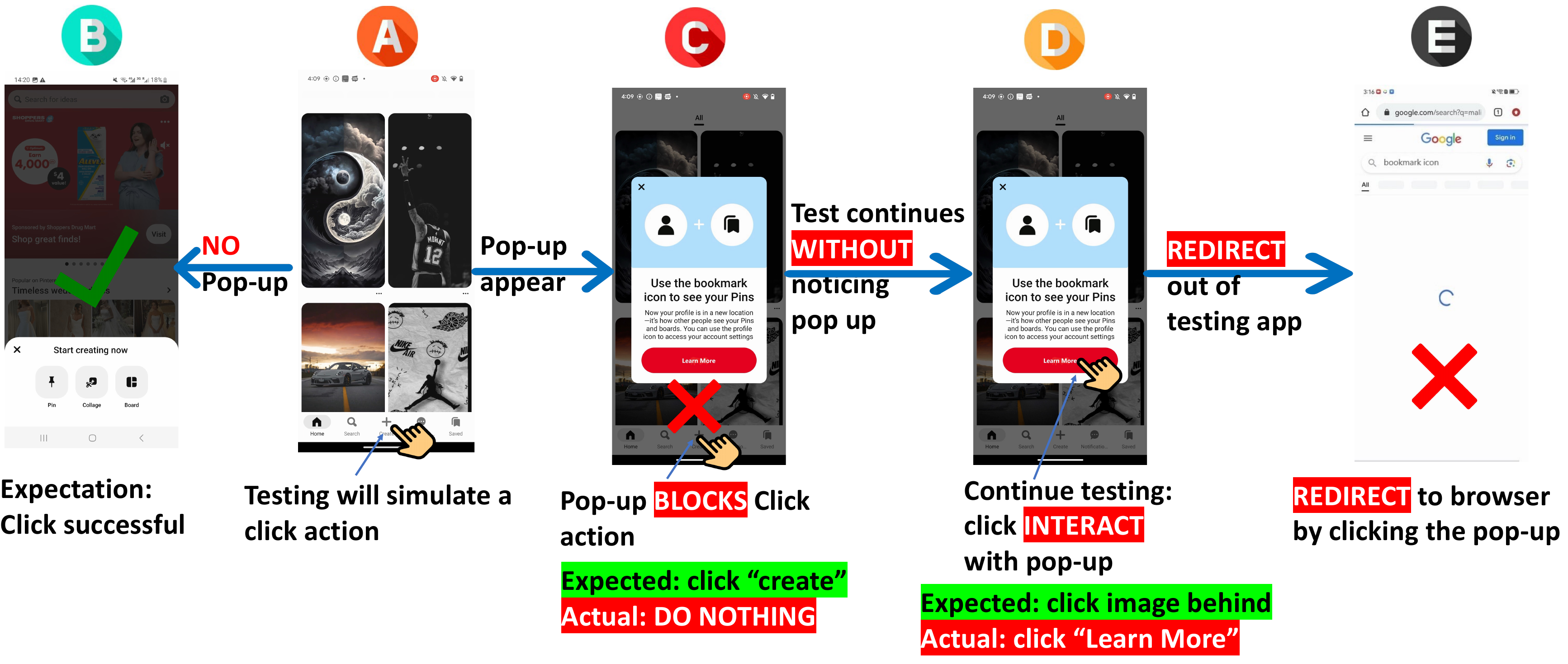}
  \vspace{-0.3cm}
	\caption{An illustrative example on the impact of app-blocking pop-ups on automated testing.}
 \vspace{-0.3cm}
	\label{fig:example_GUI_testing}
\end{figure}

However, a major challenge arises when unexpected app-blocking pop-ups, such as ads or system alerts, appear during testing. These pop-ups can obscure UI elements, disrupt the test flow, or trigger unintended actions by blocking the script’s access to the underlying interface. Since these pop-ups do not always appear and are often not accounted for in the test scripts, they can cause the tests to fail or lead to inaccurate test results. 
Figure~\ref{fig:example_GUI_testing} shows an illustrative example on the impact of these app-blocking pop-ups. 
In Figure~\ref{fig:example_GUI_testing}A, clicking the ``create'' button will open a panel for creating a post, as shown in Figure~\ref{fig:example_GUI_testing}B. However, if a pop-up appears (as shown in Figure~\ref{fig:example_GUI_testing}C), it blocks the clicking action and disrupts the test. 
In more severe cases, as shown in Figure~\ref{fig:example_GUI_testing}D, the test script may continue executing without noticing the pop-up, leading to unexpected subsequent actions interacting with the pop-up rather than the app content. This can result in undesired outcomes, such as jumping out from the current app and redirecting to a web browser, as shown in Figure~\ref{fig:example_GUI_testing}E, disrupting the automated testing process.

Existing automated testing frameworks are not generally equipped to handle these pop-ups effectively. Testers have to manually intervene, which undermines the efficiency and reliability of the testing process~\cite{2019_EMSE_Scripted_GUI_testing_of_Android_open_source_apps}. 
This issue is especially problematic for certain stakeholders. For example, phone manufacturers need to consistently test a large number of apps for comparative testing across devices and platforms. Some app developers may need to frequently compare their own apps against competitors to assess performance and usability. In such cases, app-blocking pop-ups can skew performance metrics, making it difficult to assess how well an app runs on different devices. 
This challenge becomes even more significant when developers need to continuously conduct regression testing or performance benchmarking across a wide variety of apps. Manually resolving app-blocking pop-ups significantly increases testing costs and can lead to unreliable results. 

To address this challenge, we introduce \tool, which allows real-time monitoring of an ongoing GUI test, identifies app-blocking pop-ups, and helps resolve them by providing the close button location. \tool contains three main components: 1) Real-time screenshot processing, which continuously captures screenshots during an ongoing GUI test. To improve \tool's efficiency, we apply the histogram differencing algorithm to only send screenshots that show significant changes to the next component. 2) App-blocking pop-up classifier, which is a two-stage classification pipeline that
combines two image classification models ResNet50~\cite{he2016ResNet} and MobileNetV2~\cite{sandler2018mobilenetv2} to classify if a screenshot contains app-blocking pop-ups. 3) Close button
detector, which utilizes YOLO-World~\cite{2024_CVPR_YOLO-World} to recognize close buttons with various designs, regardless of size, shape, and location. 
Finally, \tool returns the exact coordinates of a close button to the test script to trigger a click to dismiss the pop-up and allow the test to proceed smoothly. Different from most existing works that try to improve code coverage or bug reproduction
~\cite{2020_ICSE_ComboDroid_test_inputs_for_Andorid_apps_via_use_case_combinations, 2017_FSE_Guided_stochastic_model_based_GUI_testing, 2017_ICSE_DroidBot_UI_guided_test_input_generator, 2019_ICSE_APE_Practical_GUI_testing_of_Android_applications,
2022_ASE_DeepGUI_black_box_GUI_input_generation_with_deep_learning, 2020_ISSTA_Reinforcement_learning_based_curiosity_driven_testing_of_Android_applications, 2021_DeepGUIT_Deep_Reinforcement_Learning_based_GUI_Testing, 2022_TOSEM_ARES_Deep_Reinforcement_Learning_for_Black_box_Testing, 2024_ICSE_DQT_Deeply_Reinforcing_Android_GUI_Testing, 2024_ICSE_Bringing_Human_like_Interaction_to_Mobile_GUI_Testing}, \tool focuses on improving mobile app GUI testing efficiencies by resolving app-blocking pop-ups so that automated GUI tests can proceed with minimal manual intervention and interruptions.

Due to limited available data in the research community, we manually reviewed over 72K app screenshots from open-source apps in the RICO dataset~\cite{deka2017rico} to identify screenshots with app-blocking pop-ups. We further expand the data by collecting the top three apps from each category of the Google Play Store and a third-party Chinese app store, resulting in 87 apps. 
We find that app-block pop-ups are common, with over 8\% of apps in RICO and 42\% of apps in the popular apps containing at least one app-blocking pop-up. 
The evaluation on the RICO and top apps shows that \tool achieves over 91.7\% and 93.5\% in precision and recall on pop-up classification, and over 93.9\% in BoxAP and 89.2\% in recall on close button detection. The end-to-end evaluation shows that \tool can solve over 87.1\% of apps' potential blockages triggered by app-blocking pop-ups with an approximately 4s overhead for a 60-second automated GUI test recording.


In summary, the paper makes the following contributions: 

\begin{itemize}
    \item 
    We manually went over 72K GUI screenshots and identified a large-scale dataset containing over 800 app-blocking pop-ups and over 71K app content screenshots from RICO and real-world apps. This dataset provides a foundation for future research in automating the detection and classification of such pop-ups, helping to benchmark and compare different techniques in the field. We have made this dataset publicly available online~\cite{zenodo_2024_13754620}. 
    \item 
    Our empirical study reveals the widespread occurrence of app-blocking pop-ups in mobile applications, over 8\% of apps in RICO, and 42\% of apps in real-world popular apps. This insight highlights the critical need for effective solutions to address these blockages, forming the basis for our research and motivating the development of an automated resolution technique. 
    \item 
    To the best of our knowledge, this is the first study to automatically classify and detect app-blocking pop-ups based on GUI screenshots to avoid blockages for automated GUI testing. We propose a lightweight computer-vision-based approach, \tool achieves more than 91.7\% precision and 93.5\% recall in app-blocking pop-up classification while exceeding 93.9\% in BoxAP and reaching 89.2\% recall for close button detection. At the app level, \tool can solve all pop-ups in over 87.1\% of the studied apps, ensuring smooth test execution. 

    \item 
    \tool achieves significant results in classifying app-blocking pop-ups and detecting close buttons. With a performance overhead of 0.0048 seconds per frame for similarity check and an additional 0.06 seconds for classification and close button detection per selected frame during a 60-second test recording, \tool is highly efficient and well-suited for real-time automated GUI testing, ensuring swift resolution of potential blockages without compromising performance.
    \item Unlike most existing research that focuses on improving code coverage or bug reproduction, \tool targets a different but crucial challenge: improve test efficiency by resolving app-blocking pop-ups. This focus ensures automated GUI testing can proceed efficiently, reducing the need for manual intervention and improving overall efficiency.
\end{itemize}

\noindent\textbf{Paper organization.} Section~\ref{sec:motivation} discusses the motivation and empirical study results. Section~\ref{sec:methodology} presents \tool. Section~\ref{sec:evaluation} evaluates \tool. Section~\ref{sec:threats} discusses threats to validity. 
Section~\ref{sec:related} reviews related work. Finally, Section~\ref{sec:conclusion} concludes the paper.

\section{Background and A Motivation Study}
\label{sec:motivation}
In this section, we first provide the background on pop-ups and their impact on automated testing. Then, we conduct a comprehensive motivational study on their prevalence. 

\subsection{Different Types of Pop-ups}

\begin{figure}
    \centering
    \begin{subfigure}{0.3\textwidth}
        \centering
        \includegraphics[height=5cm]{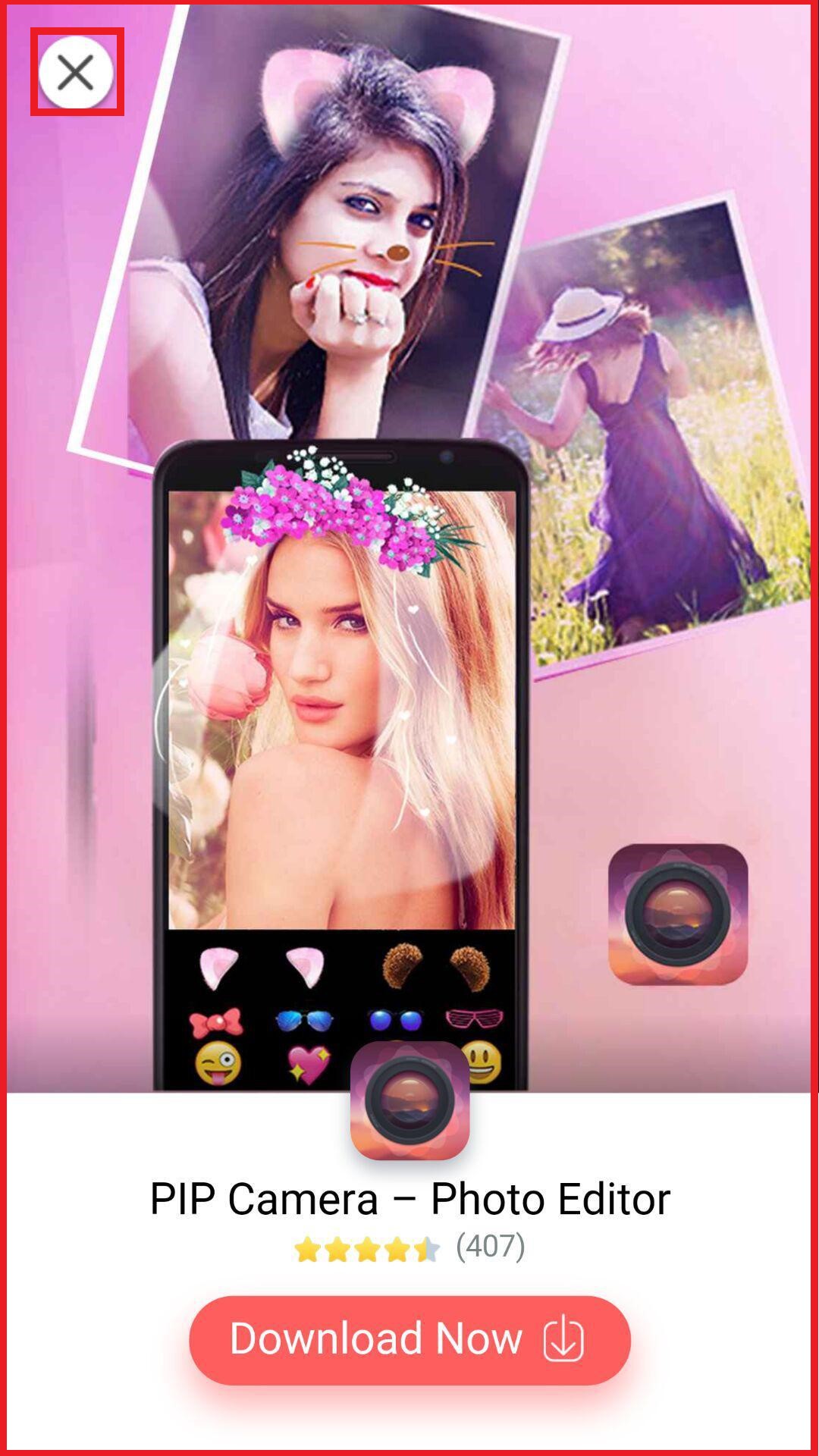}
        \caption{Fullscreen pop-up}
        \label{fig:image1}
    \end{subfigure}
    \hfill
    \begin{subfigure}{0.3\textwidth}
        \centering
        \includegraphics[height=5cm]{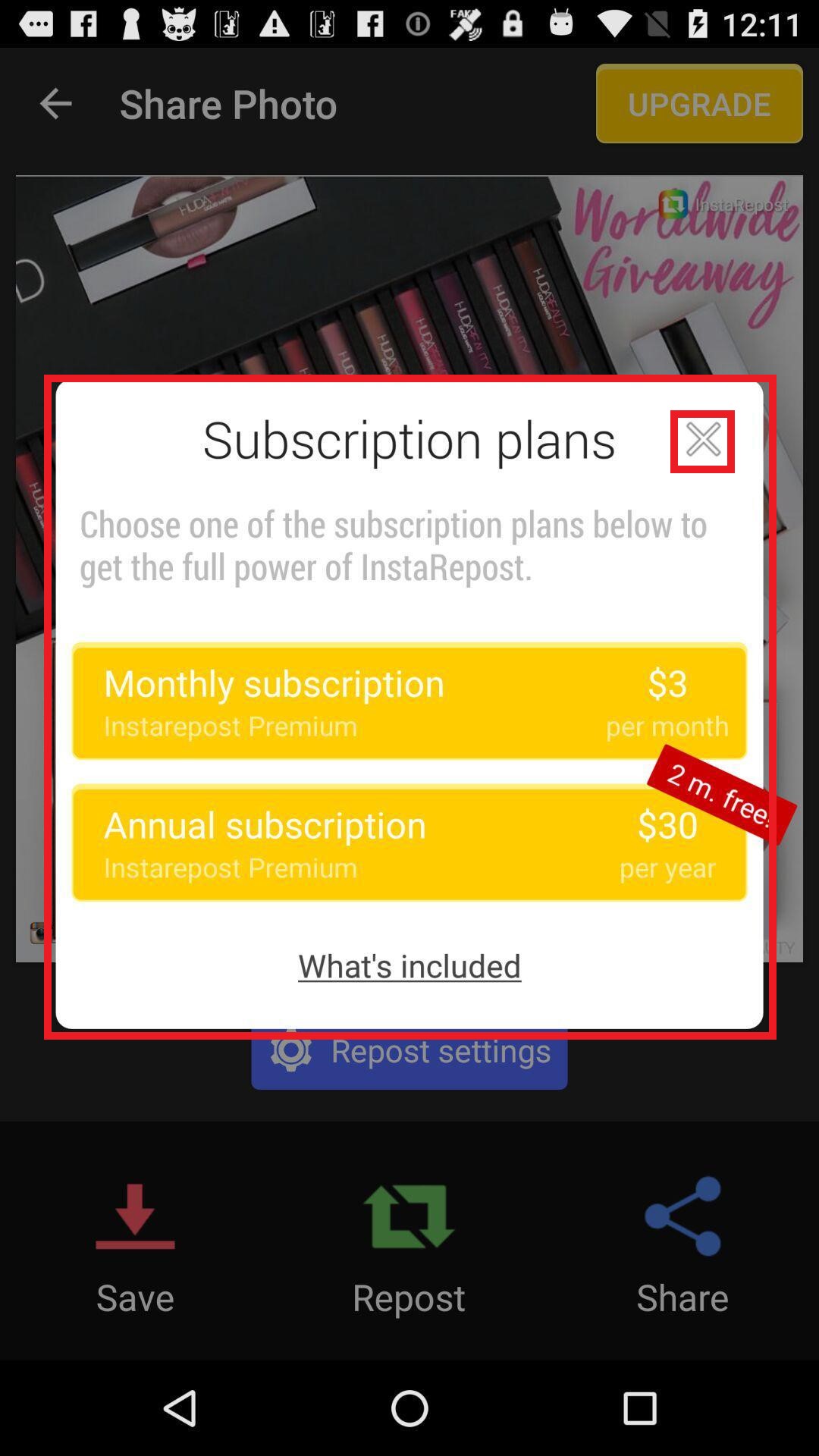}
        \caption{Partially overlay pop-up}
        \label{fig:image2}
    \end{subfigure}
    \hfill
    \begin{subfigure}{0.3\textwidth}
        \centering
        \includegraphics[height=5cm]{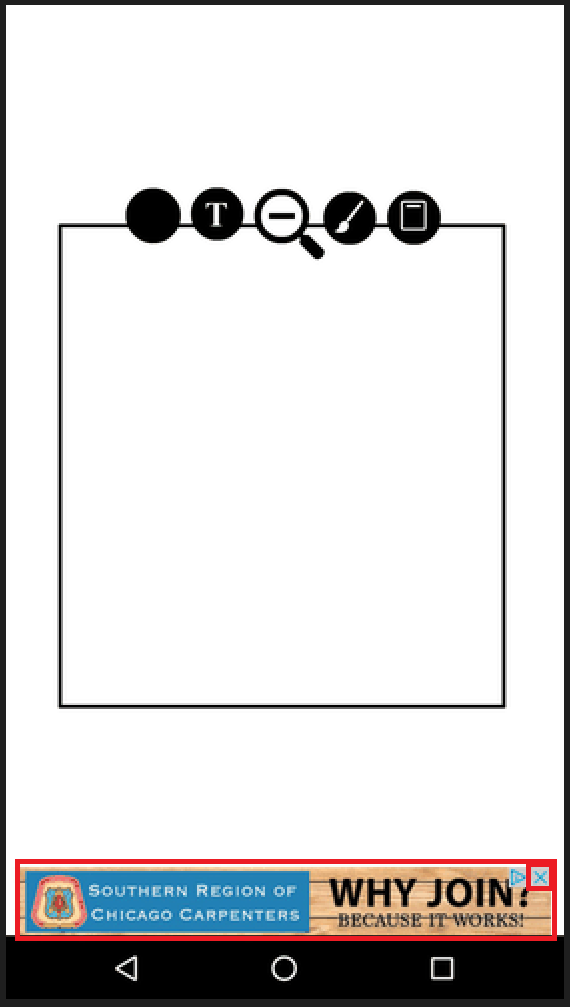}
        \caption{Banner}
        \label{fig:image3}
    \end{subfigure}
    \caption{Three types of pop-ups: full-screen, partially overlay, and banners. Red boxes mark the pop-ups and corresponding close buttons.}
    \label{fig:types_ads}
\end{figure}

In mobile applications, pop-ups are small windows or overlays that appear on top of the app's current screen (e.g., advertisement). 
Pop-ups vary in intrusiveness, size, and placement within the app’s user interface, each serving distinct purposes and presenting unique challenges. 
As shown in Figure~\ref{fig:types_ads}, there are three primary types: full-screen pop-ups, partially overlay pop-ups, and banners. 

\noindent\underline{\textit{\textbf{Full-screen pop-ups}}} occupy the entire screen of the mobile device, temporarily blocking access to the app content underneath. For example, full-screen advertisements are typically displayed at moments when the user is expected to be more receptive, such as before or after completing a task within the app. Full-screen pop-ups can include rich media content like videos or interactive elements and are often used to create a more impactful advertising experience than smaller, less intrusive formats (e.g., banner).


\noindent\underline{\textbf{\textit{Partially overlay pop-ups}}} usually appear at natural transition points within a mobile app, such as between levels in a game, during pauses, or between different content sections. These pop-ups cover a part of the screen, providing a highly engaging and immersive experience. Users must interact with or close the pop-up to return to the app content. Partially overlay pop-ups can be static images, videos, or even interactive content and are designed to capture the user’s attention by blocking the user's usage. 

\noindent\underline{\textbf{\textit{Banners}}} in mobile apps typically appear at the top/bottom of the screen or sometimes embedded within the content itself. They are smaller and less obtrusive than partially overlay or full-screen pop-ups, allowing users to continue interacting with the app while the banner content is displayed. Banners can be static or animated and usually include a clickable link that directs the user to the content website or a landing page. 

\subsection{Impact of the Pop-ups on Automated GUI Testing}
Pop-ups 
effectively capture user attention but often introduce significant challenges in automated GUI testing. These pop-ups (full-screen and partially overlay pop-ups) typically require user interaction, such as clicking a close button, before one can continue the usage process, causing significant disruption in the automated testing process~\cite{2022_IST_Understanding_app_advertising_issues_on_app_review_analysis, ruiz2014ads}. In this paper, we refer to them as \textbf{\textit{app-block pop-ups}}. 

Testers rely on automated GUI testing scripts to ensure all features function correctly and provide a smooth user experience. These scripts simulate user interactions within an app by recording and replaying various actions, such as tapping buttons, scrolling through content, and filling out forms. The scripts depend on interacting with a consistent and predictable interface to verify the app’s behavior under various conditions.
However, pop-ups often interrupt the script’s flow by introducing unexpected elements into the app’s interface. The variability in pop-up appearances—including differences in size, placement, and the presence or absence of close buttons—can render the app content non-interactive. As a result, some actions may not take effect as intended (i.e., blocked by the pop-up), and in the worst-case scenario, the testing process may get stuck. In such cases, manual intervention is required to restart the testing process, causing significant inefficiency and overhead.

In contrast, human testers can quickly adapt to these interruptions during manual testing. They identify and close pop-ups, using their judgment to navigate these obstacles. This flexibility allows them to maintain the testing flow, even with varying pop-up designs. However, automated scripts lack this adaptive capability, making handling app-blocking pop-ups a persistent challenge~\cite{giger2010predicting}. This issue underscores a critical and unresolved challenge in automated GUI testing. The current limitations of automated scripts in dealing with pop-ups highlight the need for more sophisticated solutions to ensure the reliability and effectiveness of testing processes in mobile applications.


\subsection{An Empirical Study on the Prevalence of Pop-ups}
\label{sec:empirical}

To understand the prevalence of app-blocking pop-ups in mobile applications and to investigate their potential impacts on automated GUI testing, we conducted an empirical study using data from two distinct sources. We aim to quantify how common these pop-ups are and analyze their characteristics across various apps. The findings can help better understand the challenges such app-blocking pop-ups present and work towards developing more effective solutions to improve automated app testing.

\phead{Experiment Setup. }
First, we utilized the RICO dataset
~\cite{deka2017rico}, which comprises over 72K unique UI screenshots from over 9K apps published in 2017. Although this dataset is not explicitly designed to highlight GUI pop-ups (i.e., no manual annotations), we conducted a thorough manual review of all the screenshots. This process involved meticulously examining the images to identify those containing pop-ups, and distinguishing between various types of pop-ups and other UI elements. We then documented the corresponding app names associated with these pop-ups, ensuring that our dataset accurately reflected instances of app-blocking pop-ups. This process required over 100 hours of manual effort. At first, the first author went through these UI screenshots, in cases where the classification was unclear, the second author stepped in to provide additional judgment. For instances of disagreement, the third author was involved in casting the deciding vote, ensuring consistency and accuracy across all annotations. This manual effort was crucial in building a reliable foundation for our analysis. 

Second, to expand our study beyond RICO apps, we selected three of the most popular apps from each category of the Google Play Store and a third-party Chinese app store, based on download counts, resulting in a total of 87 top-ranked Android apps. After downloading the apps, we manually completed the necessary initial setup, including permissions and introductory procedures. Then, we record 60 seconds of our manual usage of each app at a 60-frame-per-second rate. We simulated typical app usage (e.g., various clicking, scrolling, and form filling) based on the business logic during the recording. The collected data has been made publicly available to encourage further research and reproducibility~\cite{zenodo_2024_13754620}.

\begin{table}
\centering
\caption{An overview of the app-blocking pop-up distribution in studied app datasets. The `Pop-ups' column shows the number of app-blocking pop-ups in each dataset.}

\scalebox{0.9}{
\setlength{\tabcolsep}{10mm}{
\begin{tabular}{lrr}
\toprule
Dataset  & \#Apps         & \#Pop-ups  \\ \midrule
RICO              & 9K+            & 757                 \\ 
Top Google App    & 44        & 33      \\ 
Top Chinese App   & 43       & 42    \\ 
Total             & 9K+       &832                 \\ \bottomrule
\end{tabular}}}
\label{tab:popups_app_content}
\end{table}

\phead{\textit{8.3\% of RICO apps have screenshots containing app-blocking pop-ups.}} Table~\ref{tab:popups_app_content} shows the number of studied apps and app-blocking pop-ups in each data. 
Among over 72K app screenshots in RICO, we identified 757 screenshots that contain app-blocking pop-ups -- specifically, full-screen and partially overlay pop-ups. These screenshots span over 740 different apps. 
However, it’s important to note that these RICO screenshots were extracted from videos that typically capture only a single use case or interaction within an app. This means the dataset may not fully capture the extent of app-blocking pop-ups, as some may occur outside the brief interactions recorded. To address this limitation and gain a broader understanding, we extended our study to include top-ranked apps from app stores. We found that 
\textit{\textbf{42\% of the most popular apps from app stores exhibited at least one instance of app-blocking pop-ups, significantly higher than RICO apps.}} This high prevalence of app-blocking pop-ups underscores the necessity of developing robust detection and mitigation strategies in automated GUI testing. 
Approximately 75\% of the identified app-blocking pop-ups required user interactions to close, bringing significant disruptions to automatic GUI testing. The frequent occurrence of these pop-ups, especially in widely used applications, presents a significant challenge to ensuring consistent and reliable test outcomes. 

In short, our findings show that app-blocking pop-ups are common in mobile apps. Having these pop-ups can disrupt automated app testing. These disruptions are particularly costly for stakeholders such as phone manufacturers, ecosystem providers, and app developers, who often need to test multiple apps simultaneously for regression testing or performance and compatibility assessments. For example, phone manufacturers struggle to assess app performance and compatibility due to skewed metrics caused by intrusive pop-ups~\cite{cruz2019attention}. App developers and testers face challenges ensuring that apps perform well across various devices and scenarios when pop-ups interfere \cite{testing_tools2023, app_development2020}. Hence, there is a need for a non-intrusive approach that can automatically detect and close app-blocking pop-ups to ensure smooth test execution.  

\section{Approach}
\label{sec:methodology}

\begin{figure*}
  \centering
  \includegraphics[width=\linewidth]{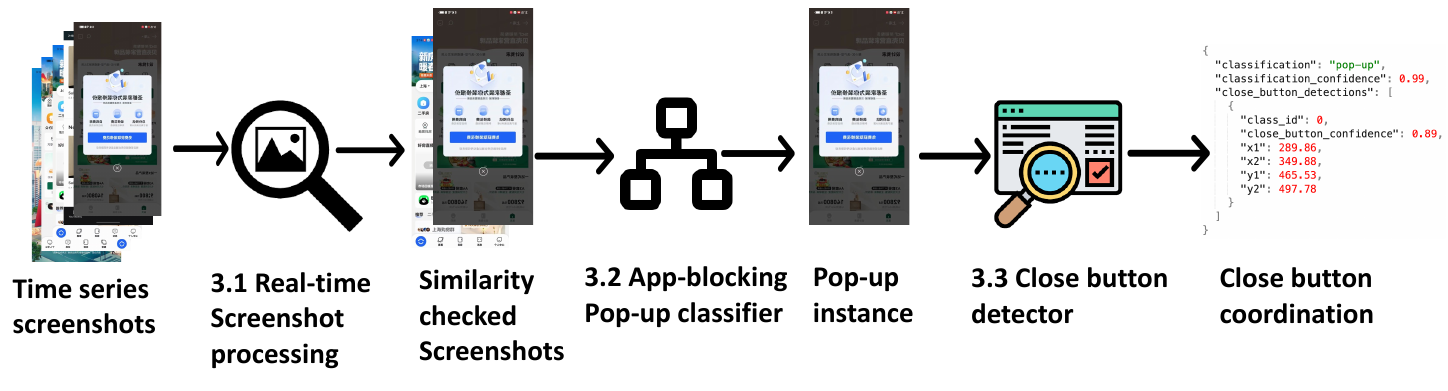}
  \caption{An overview of \tool.}
  \label{fig:illustration}
\end{figure*}

Figure \ref{fig:illustration} provides an overview of the \tool. 
\tool has three components, which collectively aim to detect and resolve GUI blockages caused by pop-ups, ensuring smooth and efficient GUI testing. (i) \underline{\textbf{\textit{Real-time screenshot processing}}}, which takes in a time series of app screenshots (e.g., streamed video during real-time GUI testing) as input. To optimize for performance and reduce overhead, we sample a screenshot every 100ms and apply image differencing algorithms to send only those screenshots with theme changes to subsequent components. 
(ii) After receiving a screenshot, \tool's \underline{\textbf{\textit{app-blocking pop-up classifier}}} uses a two-stage classification pipeline that combines the strengths of ResNet50 and MobileNetV2 models to accurately identify GUI screenshots containing pop-ups. (iii) Finally, after identifying a pop-up, \tool sends it to the \underline{\textbf{\textit{close button detector}}}, which uses the YOLO-World model~\cite{2024_CVPR_YOLO-World} to locate the close button in detected app-blocking pop-ups. 
\tool returns the exact coordinates of the button that can be passed to automation
scripts to create the necessary actions to close the pop-up and continue the GUI testing without
interruption. Below, we delve into each component in detail.

\subsection{Real-Time Screenshot Processing}

We developed \tool to efficiently analyze screenshots streamed during ongoing GUI tests in real-time to detect and close pop-up windows that obstruct the test. To meet the demands of real-time analysis, we design \tool to operate with minimal overhead, ensuring that it does not significantly impede the testing process. 
One major performance challenge is that screenshots are often captured/streamed at 60 or higher frames per second. Hence, a one-minute test will result in over 3.6K screenshots that must be analyzed, causing significant overhead. 

We use image sampling and computer vision algorithms to address the challenge of managing the large volume of time-series screenshots generated during GUI testing. First, \tool samples the transmitted screenshots at 100 millisecond intervals, balancing thorough monitoring and testing performance. 
We chose this interval because 100 milliseconds is the threshold between noticeable and unnoticeable delay for humans interacting with a UI~\cite{liu2014effects}. By sampling at this rate, \tool ensures it captures relevant GUI events in real-time without introducing significant processing overhead. 
Once the screenshots are sampled, we apply a computer vision algorithm, histogram similarity check, to determine whether each newly sampled screenshot in the time series warrants further processing. 
The histogram similarity compares the RGB histograms of consecutive screenshots to detect any significant changes in the UI. Since it compares the distributions of images, it is an efficient and accurate approach to identify significant changes in UI for real-time applications. A higher value of the histogram similarity means more similarity in the looks of the two screenshots. In this study, we use 0.8 as the similarity threshold, ensuring that unnecessary duplicate screenshots are filtered out while minimizing the risk of missing pop-up screenshots. The screenshot is then sent to the next component of \tool for an app-blocking pop-up classification.

\subsection{App-blocking Pop-up Classifier}
 Once the real-time screenshot processing component identifies a potential change in the GUI, we pass the screenshot to the pop-up classification component. This component determines whether a screenshot contains an app-blocking pop-up. We leverages a two-stage classification pipeline that combines the strengths of ResNet50 and MobileNetV2 to improve classification accuracy. 

ResNet50~\cite{he2016ResNet} and MobileNetV2~\cite{sandler2018mobilenetv2} 
are widely used in image classification due to their strong performance and efficiency. ResNet50 is a deep convolutional neural network pre-trained on the ImageNet dataset, containing over a million images classified into 1,000 categories~\cite{deng2009imagenet}. ResNet50's architecture is based on residual learning, featuring 50 layers with skip connections that help mitigate the vanishing gradient problem, making it particularly effective for training very deep networks and for capturing intricate details in images, such as textures and edges. 
MobileNetV2 is also pre-trained on ImageNet with a special focus on improving efficiency. 
It uses depth-wise separable convolutions, significantly reducing the number of parameters and computational costs compared to traditional convolutional layers. MobileNetV2 also incorporates an inverted residual structure with linear bottlenecks, which helps maintain performance while keeping the model lightweight. 
MobileNetV2 focuses more on essential, lightweight features such as edge and boundary detection, 
delivering a good balance between accuracy and computational efficiency~\cite{jaiswal2021mbnv2}. 

We implemented a sequential two-stage classification approach to combine ResNet50 and MobileNetV2. 
This strategy is inspired by techniques in other domains, such as network intrusion detection and hyperspectral image classification, where a secondary classifier refines an initial broad categorization to improve precision and reduce false positives~\cite{Sahu2015, Fouladi2016, Thaseen2016}. In our approach, ResNet50 serves as the primary classifier due to its deep architecture and strong ability to capture intricate visual patterns. If ResNet50 identifies a screenshot as containing a pop-up, the screenshot is then passed through MobileNetV2 for secondary verification. This two-stage classification leverages the computational efficiency of MobileNetV2 while relying on the superior pattern recognition capabilities of ResNet50.
This integration ensures that while the system minimizes false positives, it also avoids missing genuine pop-ups, balancing the strengths of both models to deliver reliable results in real-time applications. 

To adapt both pre-trained models for classifying screenshots containing app-blocking pop-ups, we customized their architectures by replacing the original layers with tailored heads for binary classification. For ResNet50, we introduced a custom head consisting of a linear layer, ReLU activation, dropout for regularization, and a final linear layer with a sigmoid activation, enhancing the model’s ability to distinguish between screenshots with and without pop-ups~\cite{customresnet}. Similarly, MobileNetV2 was modified with a custom head featuring a linear layer and sigmoid activation, complemented by a dropout layer. This setup preserves MobileNetV2's efficiency while improving its binary classification performance
~\cite{sandler2018mobilenetv2}. After customizing the architectures, we fine-tuned both models with GUI screenshots to further enhance their ability to detect pop-ups. Fine-tuning involved unfreezing the top layers of each model and retraining them with a lower learning rate, which allowed the models to adjust their parameters more precisely to the characteristics of our dataset~\cite{fine_tune_cnn, subeesh2021mbnv2}. For both ResNet50 and MobileNetV2, we employed the Stochastic Gradient Descent (SGD) optimizer with momentum to ensure stable updates during training. We used Binary Cross-Entropy Loss to guide the models in learning the distinctions between screenshots with and without pop-ups. 

\subsection{Close Button Detector}

Once a screenshot is classified as a potential app-blocking pop-up, we apply the close button detector. 
Since the close buttons have diverse forms (e.g., circular buttons with an ``X'',  buttons in various shapes labeled ``Close'' or in other languages, semi-transparent overlays with minimalistic designs, and more), detecting their locations can be challenging using traditional computer vision techniques. Hence, we employ the YOLO-World model~\cite{2024_CVPR_YOLO-World}, a state-of-the-art large vision model renowned for its precision and efficiency in object detection. YOLO-World’s real-time efficiency is achieved through its streamlined architecture, which allows the model to process entire images in a single pass, drastically reducing computational overhead while maintaining high accuracy. Its multi-scale feature extraction enhances its ability to detect objects of varying sizes, making it particularly effective in complex visual environments. 

We fine-tune YOLO-World to detect close buttons within GUI screenshots containing pop-ups. To achieve this, we leveraged a set of pop-up instances collected from the motivation study (i.e., RICO and popular apps that we collected), which provided a diverse range of examples with varying layouts. These instances are crucial in fine-tuning YOLO-World to focus on the unique features of close buttons within the context of mobile GUI. Meanwhile, YOLO-World's grid-based detection mechanism allows it to handle the variation in close button designs by making accurate predictions across different regions of an image~\cite{2024_CVPR_YOLO-World}. The model processes input screenshots resized to a standard dimension of 640x640 pixels and outputs the coordinates (x1, y1, x2, y2) that define the bounding box of the detected close button. 

\section{Evaluation}
\label{sec:evaluation}
In this section, we first introduce our collected data, and then we evaluate \tool by answering three research questions (RQs). 

\subsection{Data Collection and Model Training}
\label{sec:dataset}

\phead{Data Collection: }
To collect a dataset for training and evaluating our models, we utilized the open-source Rico dataset~\cite{deka2017rico} that we analyzed in Section~\ref{sec:empirical}, which comprises 72,218 mobile app screenshots from approximately 9K apps. We manually examined each screenshot to label it as either containing or not containing app-blocking pop-ups.
In total, we identified 757 screenshots containing app-blocking pop-ups across 748 unique apps and 71,461 screenshots without such pop-ups. 
For our study, we selected all 757 screenshots containing app-blocking pop-ups from 748 unique apps to train and test the computer vision models. It is important to keep a more balanced ratio~\cite{2024_Just_in_Time_crash_prediction} between the two classes (i.e., whether a screenshot contains an app-blocking pop-up) to avoid biases during training. Hence, we randomly sample 1,200 app-content screenshots from various apps for training and testing. 
Given that app-blocking pop-ups and app-content instances can originate from the same app, we ultimately selected approximately 1,000 unique apps from the Rico dataset for our study. 
In addition to utilizing the RICO dataset, we incorporated data from 87 top-ranked apps collected from app stores. To maintain balance in the dataset, we extracted 75 frames with pop-ups and 75 with app content from the top-ranked apps we collected. 

To further improve the model's generalizability, we applied oversampling by duplicating the top-ranked app data, adding 369 instances to the pop-ups class and 360 instances to the app content class. This brought the total training set to 1,994 instances (869 with pop-ups and 1,125 without), ensuring a more balanced dataset. In total, we used 1,201 screenshots containing app-blocking pop-ups (832 original + 369 oversampled) and 1,635 screenshots without app-blocking pop-ups (1,275 original + 360 oversampled) to form our dataset for training and evaluating our models. This systematic approach enhanced the model’s ability to generalize effectively without over-representing any particular source, resulting in a robust and unbiased dataset for model evaluation.

\subsection{RQ1: How accurately can \tool classify app-blocking pop-ups?}
\phead{Motivation.}
Given a mobile app GUI screenshot, \tool classifies whether or not it contains app-blocking pop-ups (i.e., \toolClassify). The classification accuracy is crucial because it affects the subsequent close button detection. 
In this RQ, we evaluate \toolClassify's app-blocking pop-up classification accuracy.


\phead{Approach.}
As detailed in the previous subsection, we utilized a curated subset of the Rico dataset and the top-ranked apps (based on downloads) we collected from app stores. This data comprises 2,836 screenshots categorized into two classes: those that contain app-blocking pop-ups and those that do not. 
We follow prior studies to split the screenshots into training, validation, and testing based on a 6:2:2 ratio~\cite{2020_FSE_Real_time_incident_prediction, 2023_TSE_App_Crowdsourced_Test_Report_Consistency_Detection}. We ensure no overlapping apps between the training and testing sets to minimize biases in the evaluation. Our training set comprises 1,994 screenshots, to which we applied a series of data augmentations and oversampling techniques to enhance model robustness. Specifically, we implemented brightness variations, contrast adjustments, darkness simulation, and Gaussian blur to introduce variability and improve generalization~\cite{yang2023imagedataaugmentationdeep}. We fine-tuned our models with our training set for 100 epochs, which took approximately 1 hour per model, using an Ubuntu server with an NVIDIA Tesla A100 GPU, an AMD EPYC 7763 64-core CPU, and 256GB RAM.


We compare \toolClassify with five baseline models, including pre-trained models, a custom convolutional neural network (CNN) model, and a large vision modeling in a zero-shot setting. These baselines were selected by their widespread use, effectiveness in image classification tasks, and unique strengths that align with our research objectives. 
Specifically, we adopted three pre-trained models, including ResNet50, MobileNetV2, and VGG19, all of which have been proven effective in a variety of image classification challenges~\cite{he2016ResNet, sandler2018mobilenetv2, simonyan2015vgg19}. These models leverage deep convolutional architectures and are trained on large-scale datasets, such as imageNet~\cite{deng2009imagenet}, enabling them to extract complex, hierarchical features from images. We fine-tuned these models using the same setting that we used for the \tool.

In addition to these pre-trained models, we developed a custom CNN, following a prior study\cite{2023_ICSE_Efficiency_Matters_Speeding_Up_Automated_Testing_with_GUI_Rendering_Inference}, for binary image classification. This model offers a more lightweight, specialized approach, with three convolutional layers followed by max-pooling, a fully connected layer, and a final sigmoid-activated output layer~\cite{oshea2015cnn}. The custom CNN was optimized using Binary Cross-Entropy loss and the Adam optimizer. 
Finally, we used a large vision model, CLIP~\cite{radford2021cliplearningtransferablevisualmodels}, for zero-shot classification, which allows the model to generalize to new classes without requiring additional training data. Comparing the classification result with CLIP provides initial evidence of the effectiveness of a large vision model on mobile-related data. 


\phead{Evaluation Metrics.}
Given that our problem is framed as an image classification task, 
we utilized three commonly used metrics i.e., precision, recall, and F1-score, to evaluate the model classification accuracy. Precision is the proportion of GUI screenshots \textit{correctly classified} as app-blocking pop-ups among all screenshots classified as app-blocking pop-ups.
\begin{equation}
\textit{Precision} = \frac{\textit{\#Screenshots correctly classified as app-blocking pop-ups}}{\textit{\#All screenshots classified as app-blocking pop-ups}}
\end{equation}
Similarly, recall is defined as: 
\begin{equation}
\textit{Recall} = \frac{\textit{\#Screenshots correctly classified as app-blocking pop-ups}}{\textit{\#All screenshots that contain app-blocking pop-ups}}
\end{equation}
Finally, the F1-score  represents the harmonic mean of precision and recall, effectively integrating both of these metrics as
\begin{equation}
\textit{F1-score} = 2 \times \frac{\textit{Precision} \times \textit{Recall}}{\textit{Precision} + \textit{Recall}}
\end{equation}

\begin{table}
\centering
\caption{Classification results of app-blocking pop-ups.}
\label{tab:results_identify_ads}
\scalebox{0.9}{
\setlength{\tabcolsep}{7mm}{
\begin{tabular}{lrrr}
    \toprule
    Method & Precision & Recall & F1  \\
    \midrule
    \textit{\textbf{Supervised learning}} & & &  \\
    CNN & 0.869 & 0.880 & 0.875  \\
    ResNet50 & 0.868 & 0.946 & 0.905  \\
    MobileNetV2 & 0.842 & 0.958 & 0.896  \\ 
    VGG19 & 0.873 & 0.841 & 0.857  \\
    \midrule
    \textit{\textbf{Zero-shot learning (Vision Model)}}& & & \\
    clip-vit-large-patch14 & 0.470 & 0.960 & 0.631 \\
    clip-vit-base-patch32 & 0.440 & 0.990 & 0.609 \\
    \midrule
    \textit{\textbf{Our approach}}& & &  \\
    \toolClassify & 0.917 & 0.935 & 0.926\\
    \bottomrule
\end{tabular}}}
\end{table}

\phead{Results. }
\textbf{\textit{\toolClassify achieves an F1 score of 0.926, providing more than 2\% improvement to the best-performing baseline model (with an F1 score of 0.905), and almost 5\% improvement on precision.
}} Table~\ref{tab:results_identify_ads} shows the classification results of \toolClassify and the baselines. 
The supervised models outperform large vision models in zero-shot learning. 
MobileNetV2 demonstrates the highest Recall at 0.958, indicating its strong ability to correctly identify app-blocking pop-ups. However, its Precision is lower at 0.842, reflecting a higher rate of false positives.  ResNet50 also shows strong Recall at 0.946 but with a lower Precision of 0.868, similar to MobileNetV2. 
\toolClassify combines ResNet50 and MobileNetV2 using model stacking, which achieves a much better precision (0.917 v.s. 0.868 and 0.842) and a similar recall (0.935 v.s. 0.946 and 0.958), achieving the highest F1 score among all models. CNN and VGG19 achieved an F1 score of 0.875 and 0.857, respectively, lower than both ResNet50 and MobileNetV2. 


The Zero-shot Learning category includes two variants of the CLIP model: clip-vit-large-patch14 and clip-vit-base-patch32. The key difference between these pre-trained models is that clip-vit-large-patch14 has a larger model size and smaller patch size, which allows it to capture more details and deliver better performance, especially in complex tasks, but at the cost of higher resource consumption.
These models show very high recalls (0.960 and 0.990, respectively) but significantly lower Precision (0.470 and 0.440), resulting in lower F1 scores (0.631 and 0.609). This suggests that while these models are highly sensitive to detecting positive instances, they also produce more false positives, reducing their overall precision. 
The finding indicates that although large vision models like CLIP are effective in broad detection tasks in zero-shot settings, they struggle with the specificity required in more targeted scenarios like pop-up detection during GUI testing. In short, supervised learning models, which are trained specifically on the task at hand, still outperform large vision models in scenarios where high precision is essential. 
Additionally, CLIP’s computational overhead is a concern for real-time applications like \tool, where processing speed is critical. Multi-modal large language models (MLLMs) like CLIP typically require more computational resources due to their larger architectures and number of parameters.


Despite the strong performance of our model, we still encountered misclassifications in some GUI screenshots. After manually reviewing 25 misclassified cases (11 pop-up misclassified and 14 for app content), we identified two primary reasons for these errors. The first reason is related to having complex visual information in the screenshot. For example, as shown in Figure~\ref{fig:image1}, 
the transparent pop-up overlays the app interface, making it hard to separate the pop-up from the app content. This visual overlap causes the model to misclassify it as part of the app due to the lack of clear distinction. The subtle integration of the pop-up within the app’s layout further reduces the visual contrast, making it even more difficult for the model to detect the pop-up.
The second reason is that the design of the app content can be ambiguous. For example, in Figure~\ref{fig:image2}, the app content has a design with square boxes with ``X'' shape close buttons (marked in red). Such elements are often used in pop-ups, confusing the model, and leading to misclassification. 
Although \toolClassify achieves high classification accuracy, our results also find that app design has a wide variation, which can result in misclassification. Future studies should consider increasing the training dataset to further enhance the result. 

\begin{figure}
    \centering
    \begin{subfigure}{0.45\textwidth}
        \centering
        \includegraphics[height=5cm]{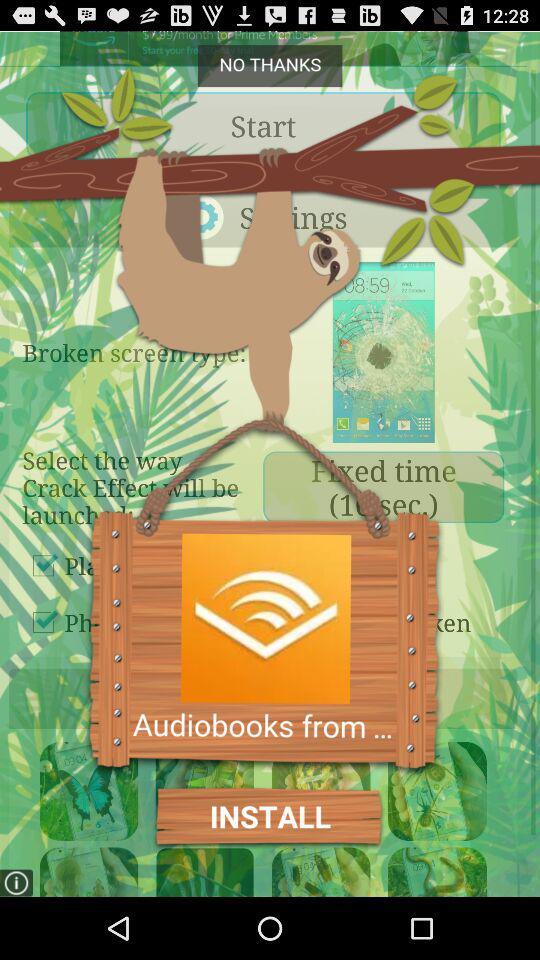}
        \caption{A pop-up instance misclassified as app content.}
        \label{fig:image1}
    \end{subfigure}
    \hfill
    \begin{subfigure}{0.45\textwidth}
        \centering
        \includegraphics[height=5cm]{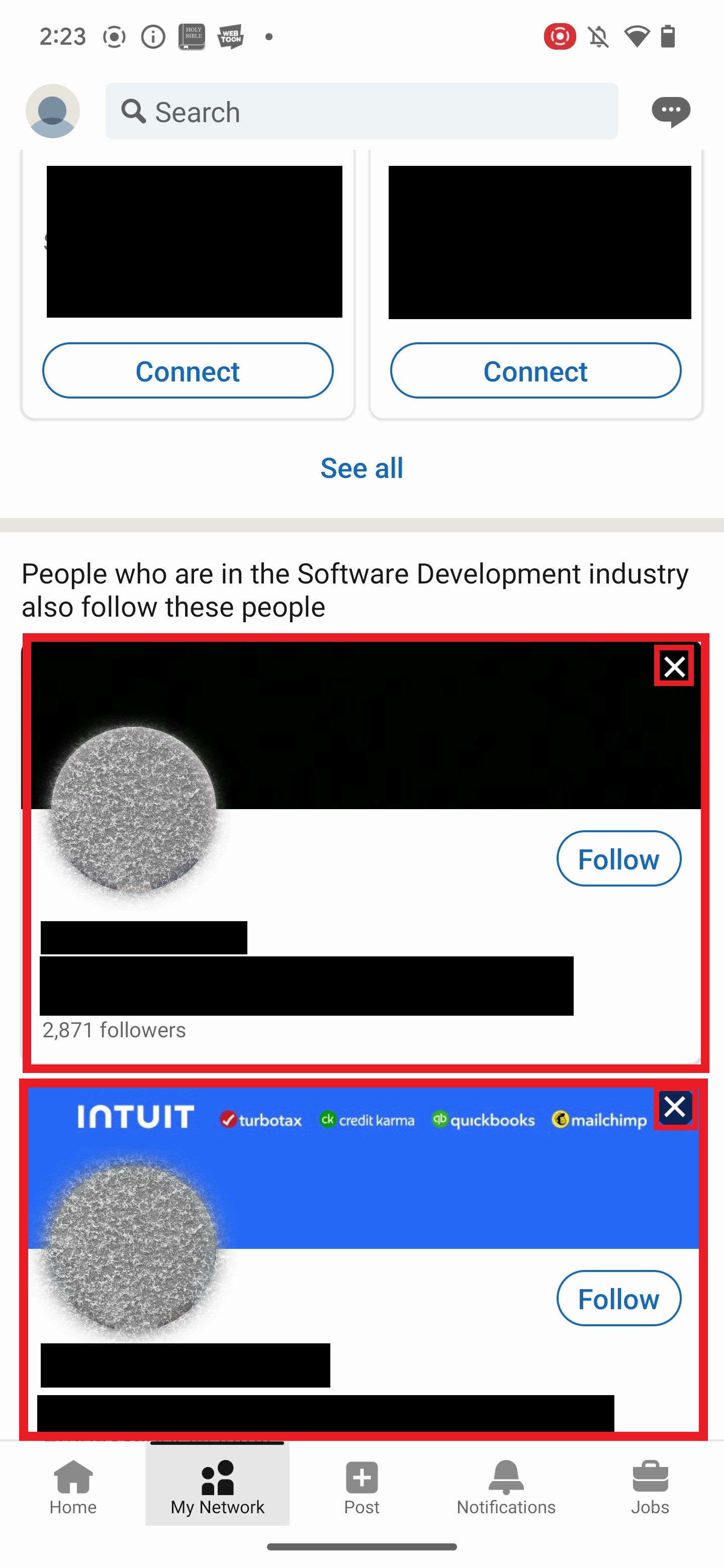}
        \caption{An app content instance misclassified as a pop-up. }
        \label{fig:image2}
    \end{subfigure}
    \caption{Examples of app-blocking pop-up misclassification.}
    \label{fig:misclassification_examples}
\end{figure}

\rqboxc{\toolClassify achieves superior performance compared to baseline models, achieving the highest F1 score of 0.926. It consistently detects app-blocking pop-ups across a wide range of app designs, demonstrating its reliability in varied and complex scenarios. }

\subsection{RQ2: How accurately can \tool detect buttons that can close app-blocking pop-ups?}

\phead{Motivation:}
In this research question, we evaluate \tool's detection and localization of close buttons in app-blocking pop-ups (i.e., \toolDetect). Accurately detecting and locating the close button is crucial for resolving interruptions caused by pop-ups in automated GUI testing.



\phead{Approach.} We fine-tuned YOLO-World~\cite{2024_CVPR_YOLO-World} using our dataset, which consists of 832 manually labeled screenshots containing app-blocking pop-ups from RQ1.
Note that \tool operates in two sequential phases: pop-up classification and button detection. The second phase ideally only receives screenshots that contain pop-ups. 
Therefore, in this RQ, we evaluate \toolDetect separately by focusing solely on screenshots with app-block pop-ups, and RQ3 contains the end-to-end evaluation result. 

We divide the dataset into training, validation, and testing sets by the 6:2:2 ratio~\cite{2020_FSE_Real_time_incident_prediction, 2023_TSE_App_Crowdsourced_Test_Report_Consistency_Detection}. To enhance the performance and robustness of our models, we applied data augmentation and oversampling techniques across the entire training set. We generated three augmented variants for each of the 490 original training images and applied oversampling by duplicating the
top-ranked app data, resulting in a total of 1,578 images for training. The augmentations included hue adjustments (ranging from -100° to +100°), saturation adjustments (from -70\% to +70\%), brightness adjustments (from -40\% to +40\%), and the addition of noise, where up to 2\% of pixels were altered to simulate realistic noise. This process introduced greater variability in the dataset, allowing the model to generalize better across different conditions. 
We apply 100 epochs in the fine-tuning process, 
with an image size of 640 pixels to match the input size of YOLO-World. The process took approximately 1 hour with using the same equipment we had in RQ1. 
The models' output provides a bounding box around the close button, represented by the coordinates (x1, y1) for the top-left corner and (x2, y2) for the bottom-right corner.

We compared \toolDetect with two other baselines: Faster R-CNN~\cite{ren2016fasterrcnnrealtimeobject} and YOLOv8~\cite{yolov8}. Faster R-CNN, or Region-based Convolutional Neural Network, is a two-stage object detection model that first generates region proposals and then applies a CNN to each proposed region to classify and refine the bounding boxes. This model is known for its high accuracy, which is crucial for tasks requiring precise localization of small objects like close buttons, but with higher computation overheads. 
YOLOv8 is distinguished by its real-time object detection capabilities, balancing speed, and accuracy. By dividing the image into grids and predicting bounding boxes and class probabilities simultaneously, YOLOv8 excels in scenarios demanding quick processing without compromising precision, making it ideal for detecting close buttons in real-time applications.

\phead{Evaluation Metrics.} 
To evaluate our models' performance in accurately detecting and localizing close buttons in app-blocking pop-ups, we selected a set of commonly-used metrics in objection detection -- Recall, BoxAP (Bounding Box Average Precision), mAP@50 (Mean Average Precision at the Intersection over Union, IoU, of 50\%), and mAP@50-95 (Mean Average Precision across multiple IoU thresholds). These metrics provide a comprehensive assessment of the models' precision and robustness in various detection scenarios~\cite{kdnuggetseval2022MAP, yolov8heritage2024MAP}. 
First, Recall represents the ratio of correctly identified close button bounding boxes to all actual close button instances in the ground truth:
\begin{equation}
\textit{Recall} = \frac{\text{True Positive Bounding Boxes}}{\text{True Positive Bounding Boxes} + \text{Missed Close Buttons (False Negatives)}}.
\end{equation}

Precision is defined as:
\begin{equation}
\textit{Precision} = \frac{\text{True Positive Bounding Boxes}}{\text{True Positive Bounding Boxes} + \text{False Positive Bounding Boxes}},
\end{equation}
which calculates the ratio of correctly identified close button (true positives) to the total identified close button (true positives + false positives). However, in object detection, simply determining if an object is correctly identified is not enough -- it is crucial to measure how accurately the bounding box matches the true object location. Thus, we use BoxAP to measure the accuracy of the bounding box predictions, which measures how accurately the model detects the bounding box around the close button. We defined BoxAP as:
\begin{equation}
\textit{BoxAP} = \frac{1}{N} \sum_{i=1}^{N} \text{Precision}(i) \times \Delta \text{Recall}(i),
\end{equation}
\text{where } N \text{ is the number of bounding boxes, } \text{Precision}(i) \text{ is the precision at each threshold } i, and \text{$\Delta$Recall}(i) \text{ is the change in recall between threshold } i \text{ and } i-1.

We also report the mAP@50 and mAP@50-95 by calculating the mean average precision across different percentages of IoU. The value of IoU is in ranges from 0 to 100\%, where 100\% is a perfect match. For instance, mAP@50 measures the mean of AP values across all detected close buttons at an IoU threshold of 50\%. IoU is defined as: 
\begin{equation}
\textit{IoU} = \frac{\text{Area of Overlap between Predicted and Ground Truth Close Button Boxes}}{\text{Total Area Covered by both Predicted and Ground Truth Boxes}}.
\end{equation}
Subsequently, mAP@50 and mAP@50-95 are calculated as:
\begin{equation}
\textit{mAP@50} = \frac{1}{N} \sum_{i=1}^{N} \text{AP}_i^{\text{IoU} = 0.5}
\end{equation}
\begin{equation}
\textit{mAP@50-95} = \frac{1}{10} \sum_{k=0}^{9} \left( \frac{1}{N} \sum_{i=1}^{N} \text{AP}_i^{\text{IoU} = 0.5 + 0.05k} \right)
\end{equation}


\begin{table}
\centering
\caption{Results of locating closing buttons of app-blocking pop-ups. Metrics are calculated on the test dataset.}
\label{tab:results_locating_ads}
\scalebox{0.9}{
\setlength{\tabcolsep}{6mm}{
\begin{tabular}{lrrrr}
    \toprule
    Method & BoxAP & Recall & mAP50 & mAP50-95  \\
    \midrule
          Faster R-CNN&  0.943  &  0.864 &    0.930 &  0.739\\
          YOLOv8&     0.939     &   0.892     &  0.918  & 0.758 \\
          \toolDetect (YOLO-World)& 0.939      &   0.892     &  0.936  &  0.760 \\
    \bottomrule
\end{tabular}}}
\end{table}

\phead{Results. }\textit{\textbf{\toolDetect(YOLO-World) achieves a Recall of 0.892 and a mAP@50 score of 0.936, indicating its strong ability to detect and accurately localize close buttons.}} Table~\ref{tab:results_locating_ads} shows the close button detection result. Compared to Faster R-CNN, which has a lower Recall of 0.864, YOLO-World demonstrates a 3.24\% improvement in recall, meaning that it captures more relevant instances during detection. Additionally, YOLO-World outperforms Faster R-CNN in mAP@50 by 0.65\%, with a score of 0.936 compared to 0.930, highlighting its superior precision at the 50\% overlap threshold. 
Although YOLOv8 achieves the same Recall as YOLO-World, it has a slightly lower mAP50 and mAP50-95. 
This combination of YOLO-World's higher recall and better mAP@50 underscores its robustness in scenarios requiring both comprehensive detection and precise localization, making it a more reliable model for \toolDetect.

While \toolDetect performs effectively across the testing set, 
it still misses some close buttons or detects incorrect ones. To understand the causes, we conducted further analysis. 
We found that the Rico dataset achieved higher precision and recall (96\% and 91\%, respectively) compared to the top-ranked apps (86\% and 71\%, respectively). This could be due to the consistency in the appearance of close buttons within the Rico data, where many buttons follow similar design patterns, making them easier for the model to detect accurately. However, for the top-ranked apps, precision and recall were notably lower. We examined the composition of the top-ranked apps in the testing set and found that they could be divided into two categories by app-blocking pop-ups category: apps with fullscreen pop-ups and partially overlay pop-ups. \toolDetect performed well on partially overlay pop-up apps (precision: 90\%, recall: 81.8\%), where close buttons follow more standardized formats.
In contrast, apps with fullscreen pop-ups posed a challenge(precision: 75\%, recall: 37.5\% ), as many of their close buttons are text-driven or have unique formats that differ significantly from those seen in the top-ranked apps or Rico dataset. The limited training data for these varied formats in fullscreen pop-up apps from top-rank apps likely contributed to the lower performance in detecting and accurately localizing close buttons. This lack of training data negatively impacted the overall precision and recall, lowering the model’s performance across the full testing set.

\rqboxc{With a recall of 89.2\% and mAP@50 score of 93.6\%, \toolDetect effectively detects close buttons in various app-blocking pop-up scenarios. These results show that \tool effectively resolves test disruptions caused by pop-ups.}

\subsection{RQ3: How useful is \tool when integrated into real-world automated testing scenarios?}

\phead{Motivation.}
In this RQ, we conduct an end-to-end evaluation of \tool. We investigate whether \tool can detect and resolve app-blocking pop-ups reliably, 
allowing tests to continue as intended without interruption. 

\phead{Approach.}
We conducted two experiments to assess \tool's effectiveness. The \textit{\textbf{first experiment}} is an end-to-end evaluation using models that we trained and the testing dataset from RQ1 and RQ2. 
This testing dataset comprised 167 screenshots from 154 unique apps categorized as \textit{app-blocking pop-ups} and 255 screenshots from 150 unique apps categorized as \textit{app content}. 
Since RICO's data is in GIF, we conducted a simulation-based experiment. We analyze the GIFs as individual screenshots (i.e., like a video) and feed the screenshots to \tool for sampling, app-blocking pop-up classification, and close button detection.


Due to the restriction of RICO data (i.e., only GIF available) and limited data on popular apps, we expand our data to conduct the \textit{\textbf{second experiment}} with an \textit{additional} dataset of 51 top-ranked apps, we manually collected the 4th and 5th most popular apps (The top 3 apps have already been included in Section~\ref{sec:motivation}) in the app stores by following the same steps described in Section~\ref{sec:motivation}, which resulted in 51 apps.
Because of the extensive manual overhead of writing tests for a diverse number of apps, we conduct this second experiment in a simulated fashion.
First, we created a one-minute-long usage recording for these 51 apps. 
Among these recorded usages, 25 apps contained at least one app-blocking pop-up. We annotated the pop-ups for each recording as the ground truth to evaluate \tool.
Then, we replayed the recordings of the 51 apps, sequentially sending individual frames as input to \tool according to a fixed interval (i.e., 100ms). 
If \tool detects the current frame is similar to the previous frame, we continue the simulation. Otherwise, the frame then goes through ads classification and close button detection. If \tool classifies the frame as containing an app-blocking pop-up and detects a close button, it returns the coordinates of the button in the frame. 
Finally, we compare the returned coordinates with the ground truth to evaluate the detection accuracy.

Although we conducted the experiment in a simulated fashion, \tool can be seamlessly integrated into existing automated testing workflows, enhancing efficiency and reliability. 
Testers can integrate \tool into their automated testing frameworks by embedding logic in their test scripts to periodically capture and send real-time GUI screenshots. Upon receiving a screenshot, \tool returns ``continue'' if the frame is similar to the previous frame (i.e., no change) or the close button coordinates (i.e., it detects the button in an app-blocking popup). The added logic in the test code can then perform a button click action if \tool returns a coordinate. 
By doing so, testing workflows can continue uninterrupted, minimizing required code changes to adapt.

\phead{Evaluation Metrics.} We used precision and recall to evaluate \tool’s performance for both experiments, we also used responding time for the second experiment to evaluate our tool's real-time ability. Precision reflects the proportion of correctly identified pop-ups and close buttons among all screenshots flagged as such:
\begin{equation}
\textit{Precision} = \frac{\text{Correctly classified pop-ups and close button}}{\text{All classified as app-blocked pop-ups}}.
\end{equation}

Recall measures the proportion of actual pop-ups and corresponding close buttons that were correctly detected:
\begin{equation}
\textit{Recall} = \frac{\text{Correctly classified pop-ups and close button}}{\text{All ground truth app-blocking pop-ups}}.
\end{equation}
Additionally, we measured the \tool's response time to ensure its real-time capability in handling pop-up detection and providing close button information for testers.

\begin{table}
\centering
\caption{The precision and recall of successfully identifying the close buttons in end-to-end settings.}
\label{tab:results_rq3}
\scalebox{0.9}{
\setlength{\tabcolsep}{4.5mm}{
\begin{tabular}{lrrr}
    \toprule
    Experiment & Precision & Recall & F1  \\
    \midrule
          Simulated end-to-end evaluation using RICO and top 3 apps& 0.887      &   0.844     &  0.865 \\
          Automated end-to-end evaluation using \textit{additional} top apps  & 0.871      &   0.794     &  0.831 \\
    \bottomrule
\end{tabular}}}
\end{table}

\phead{Results. }
\textbf{Our simulated end-to-end experiment shows that \tool has a precision of 88.7\% and a recall of 84.4\% on identifying app-blocking pop-ups and locating corresponding close buttons, effectively resolving the issue of app-blocking pop-ups.} Table ~\ref{tab:results_rq3} shows the precision and recall of the two end-to-end experiments. 
In the first experiment (on RICO and top three apps), \tool is able to accurately detect close buttons in 141 out of 167 app-blocking pop-ups (a recall of 84.4\%). 
\tool also has a high precision in detecting the close buttons accurately (88.7\%). 
Since an app may have multiple pop-ups, we also study if \tool can identify and resolve all pop-ups in an app. At the app level, \tool can detect and resolves all the pop-ups in 135 out of 155 apps (87.1\%), ensuring smooth test execution. 
While \tool demonstrated strong performance, we want to conduct a detailed error analysis to further identify areas for improvement. 
We find that most issues were due to misclassification, primarily because the design of these pop-ups was very similar to app content. For example, some screenshots have clean layouts where pop-ups blended seamlessly into the overall interface. These layouts often lacked the typical visual cues, such as distinct borders or shading, making it difficult for \tool to distinguish between pop-ups and regular app content. 

In the second experiment (4th and 5th most popular apps based on automated tests), \textbf{\tool demonstrated strong performance, achieving a precision of 87.1\% and a recall of 79.4\% on correctly identifying app-blocking pop-ups and locating their close buttons.} The approach detected 31 of the 34 ground-truth app-blocking pop-ups and successfully located the close buttons for 27 of them. At the app level, \tool detected and resolved app-blocking pop-ups in 19/25 apps. 
After some investigation, we find that the slight decrease in precision and recall compared to the first experiment is due to the complexity and variability of real-world close button designs. For example, close buttons can take various forms, such as unconventional shapes, semi-transparent icons, or being embedded within images that blend into the background, making them visually challenging to detect. Moreover, there are some text-based close buttons, whose text can be in either English or Chinese and can have different words (e.g., ``Close'' v.s. ``Skip''), which further complicates the detection. Moreover, since most training data in \tool is from RICO (whose apps are mainly in English), \tool achieves lower performance in Chinese apps, which have a higher presence in the second experiment. Nevertheless,  \tool still maintained high performance, showcasing its robustness in handling diverse designs and achieving reliable results.


\noindent\textbf{\tool is highly efficient, achieving an average response time of 60 milliseconds per pop-up classification and close button localization.} We conducted the performance analysis using the apps that we used for the second experiment, using the same hardware that we used for training the models. After passing the similarity check which cost approximately 0.0048 seconds per frame, the average classification time per screenshot was 0.04 seconds. For the screenshots classified as app-block pop-ups, the close button detector took an average of 0.02 seconds per screenshot (around 0.24 seconds per each one-minute test). 
\tool’s high efficiency demonstrates its practical value to help improve testing in real-time. 

While \tool demonstrates strong performance in detecting and resolving app-blocking pop-ups, it is important to analyze the error types and their impacts. These errors can be categorized into three main types. The first type occurs when app-blocking pop-ups are misclassified as app content, resulting in no returned coordinates for the close button. Although these cases account for 42\% of errors, they behave similarly to the scenario where \tool is not used, resulting in no additional disruption. 
The second type involves app contents being mistakenly identified as containing app-blocking pop-ups, resulting in either no returned coordinates or the return of invalid close button coordinates.
These errors represent 15\% of all cases, with most cases resulting in no returned coordinates, thereby avoiding any interaction in the testing workflow. For the remaining errors of this type, such as the example in Figure~\ref{fig:misclassification_examples}(b), invalid coordinates are returned. Clicking on these coordinates may result in unpredictable behavior, such as unintended interactions with the app content. However, this scenario is quite rare, accounting for approximately 10\% of cases within the second error type and approximately 1.5\% of all error cases.
The third type is when app-blocking pop-ups are correctly identified, but the returned close button coordinates are inaccurate. These errors comprise 43\% of error cases. Among these errors, 75\% of the invalid coordinates point to areas outside the bounds of the app-blocking pop-up. These interactions are unlikely to disrupt app functionality (i.e., the click action is blocked by app-blocking pop-ups), allowing the test to proceed without issue. The remaining 25\% cases of this type of error involve coordinates located within the bounds of the pop-up, which may result in unintended behavior, such as app redirection. As we mentioned in Section~\ref{sec:intro}, this challenge is not unique to \tool; existing testing workflows without the tool also face the same challenge. Both cases, whether occurring in workflows with or without \tool, may result in the test getting stuck or redirected, requiring manual intervention. 
However, the presence of \tool significantly reduces the likelihood of such situations occurring, thereby improving the overall stability and efficiency of the testing workflow.

These findings suggest that, despite the inherent challenges associated with these error cases, \tool maintains a high degree of reliability and efficiency in detecting and resolving app-blocking pop-ups. Its real-time operation and ability to adapt to diverse and complex application environments demonstrate its potential as a valuable tool for improving the robustness and efficiency of automated testing workflows.


\rqboxc{With a precision of 88.7\% and recall of 84.4\% in the simulated end-to-end evaluation and a precision of 87.1\% and recall of 79.4\% in tests that exclusively cover popular apps on app stores, \tool effectively classifies pop-ups and detects corresponding close buttons in various app-blocking pop-up scenarios. \tool is also efficient where it takes only 60ms on average to process a screenshot. }



\section{Threats to Validity}
\label{sec:threats}

\subsection{External Validity}
One potential threat to external validity is the generalizability of the dataset used in the experiment. While we try to create a diverse set by considering RICO and popular apps from two app stores, 
the dataset may still not comprehensively represent all mobile application environments. For instance, certain types of mobile ads or specific app interactions may not be captured in the dataset, leading to potential gaps when generalizing the results across other applications. Future studies should consider extending \tool to include data from more apps. 

\subsection{Construct Validity}
Construct validity may be affected by how app-blocking pop-ups and close buttons are defined and labeled in our dataset. While the RICO dataset provided a strong foundation, challenges arose in the detection of more complex and unconventional designs, such as text-based or semi-transparent close buttons in some top-ranked apps. The reliance on manual labeling for certain pop-up instances may introduce subjectivity, potentially impacting the precision of the classification and detection tasks. In particular, inconsistencies in identifying non-standard close buttons could skew the accuracy of performance metrics. 
The end-to-end app testing scenarios in RQ3 may not fully capture the real-world behavior of pop-ups. In particular, \tool's performance in detecting pop-ups and close buttons (with an average response time of 60 milliseconds) might be optimal in controlled environments but could vary under different network conditions, app behaviors, or interaction patterns. Nevertheless, we evaluated \tool on a wide range of open source and most popular apps from two app stores to minimize the threats. 

\subsection{Internal Validity}
Variations in the appearance of pop-ups and close buttons could influence the internal validity of the results. For instance, while \tool achieved a high recall of 92.4\% in identifying pop-ups in the controlled environment, real-world complexities, such as ads that auto-close or dynamically resize, may not have been fully represented in the training process. Additionally, the overfitting risk associated with data augmentation techniques (e.g., brightness and contrast changes) might inflate the model’s performance on the testing data but reduce its effectiveness in actual applications.

\section{Related work}
\label{sec:related}
In this section, we review related work  in four areas: mobile advertisements, mobile GUI exploration and testing, and using computer vision to assist mobile testing.

\subsection{Mobile Advertisements} 
Advertisements, or ads, are ubiquitous in mobile apps. 
However, having ads may also lead to various issues and bad user impressions. \citet{2015_ICSE_Hidden_Cost_of_Mobile_Ads_for_Software_Developers} found that mobile ads significantly increase network data usage and energy consumption, require frequent code changes, and lead to many user complaints, negatively impacting app ratings. 
\citet{2022_IST_Understanding_app_advertising_issues_on_app_review_analysis} analyzed ad-related user reviews and discovered that users often complain about seeing too many ads. 
\citet{2021_IST_performance_costs_of_in_app_ads_on_user_experience} also found that users are particularly concerned about the battery consumption caused by mobile ads.  
Some ad libraries 
may collect private information~\cite{2012_Unsafe_exposure_analysis_ads, 2015_arxiv_Mobile_Ad_Targeting} or request excessive permissions~\cite{2013_arxiv_Analysis_of_Android_Ad_Library_Permissions}, posing privacy risks to users. 
Prior researches also work on detecting fraudulent ads~\cite{2018_FSE_FraudDroid, 2019_WWW_MAdLife, 2020_WWW_MadDroid, 2023_ICSE_AdHere}. These works rely on analyzing Android apps' layout files~\cite{2018_FSE_FraudDroid, 2023_ICSE_AdHere} or monitoring network requests to specific third-party domains to detect ads~\cite{2019_WWW_MAdLife, 2020_WWW_MadDroid}. 

Our work focuses on a different issue that ads may bring -- blocking automated GUI tests. Due to the wide variety of apps and ad libraries, it is challenging to apply prior techniques to a large set of apps. Our approach uses computer vision techniques to analyze GUI screenshots, which do not need any app instrumentation and can be adapted easily. 

\subsection{Mobile GUI Exploration and Testing}
In mobile testing, most approaches focus on GUI-based testing. 
They often rely on GUI exploration strategies like random-based~\cite{Monkey, 2013_FSE_Dynodroid}, model-based~\cite{2017_FSE_Guided_stochastic_model_based_GUI_testing, 2017_ICSE_DroidBot_UI_guided_test_input_generator, 2019_ICSE_APE_Practical_GUI_testing_of_Android_applications, 2020_ICSE_ComboDroid_test_inputs_for_Andorid_apps_via_use_case_combinations}, and learning-based approaches~\cite{ 2022_ASE_DeepGUI_black_box_GUI_input_generation_with_deep_learning, 2020_ISSTA_Reinforcement_learning_based_curiosity_driven_testing_of_Android_applications, 2021_DeepGUIT_Deep_Reinforcement_Learning_based_GUI_Testing, 2022_TOSEM_ARES_Deep_Reinforcement_Learning_for_Black_box_Testing, 2024_ICSE_DQT_Deeply_Reinforcing_Android_GUI_Testing}.
However, these popular approaches primarily focus on exploration strategies to improve coverage and often neglect the challenge of pop-ups, which are common in mobile apps.
As a result, GUI testing with these approaches still encounters interruptions caused by pop-ups. For example, a user event intended for a specific widget (e.g., clicking a save button) may be blocked by an unexpected pop-up overlay, disrupting the testing workflow. 
Some recent studies leveraged Large Language Models (LLMs) for automated GUI testing. Liu et al.~\cite{2024_ICSE_Bringing_Human_like_Interaction_to_Mobile_GUI_Testing} use GUI page information as a prompt and feeds it to an LLM to generate test cases. Some studies~\cite{2023_arxiv_AppAgent_Multimodal_Agents_as_Smartphone_Users, 2024_arxiv_Mobile_Agent} utilized multi-modal agents to interact with mobile applications by mimicking human-like interactions. However, these multi-modal models (e.g., CLIP~\cite{radford2021cliplearningtransferablevisualmodels}) lack the ability to specifically handle pop-ups. They analyze each GUI screenshot, regardless of whether it contains pop-ups, and infer subsequent events (e.g., clicks) based on the provided prompt. Consequently, if a GUI screenshot contains a pop-up, the agents may still execute the generated events, potentially leading to disruptions caused by unexpected pop-up overlays. 
Our approach, which detects and resolves pop-ups, can complement these existing testing approaches by ensuring that GUI testing continues as expected.

\subsection{Using Computer Vision To Assist Mobile Testing}
Some studies utilized computer vision techniques for GUI analysis to support testing by improving exploration efficiency ~\cite{2023_ICSE_Efficiency_Matters_Speeding_Up_Automated_Testing_with_GUI_Rendering_Inference}, generating appropriate test input texts~\cite{2023_FSE_Appaction, 2023_ICSE_Fill_in_the_Blank_Text_Input_Generation_for_Mobile_GUI_Testing, 2024_ICSE_Unusual_Text_Inputs_Generation_for_Mobile_App_Crash_Detection_with_LLM}, recording and replaying tests~\cite{2020_ICSE_translating_video_recordings_of_mobile_app_usages, 2023_UIST_Video2Action}, and enabling cross-platform testing~\cite{2021_ICSE_Layout_and_Image_Recognition_Driving_Mobile_Testing, 2024_TOSEM_PIRLTEST_GUI_Testing_via_Image_Embedding_and_RL}. 
Feng et al.~ \cite{2023_ICSE_Efficiency_Matters_Speeding_Up_Automated_Testing_with_GUI_Rendering_Inference} use a convolutional neural network (CNN) to infer the state (i.e., fully or partially rendered) from GUI screenshots, executing the next event only when the GUI is fully rendered, thus reducing waiting time between events.
To generate valid input texts, Hu et al.~\cite{2023_FSE_Appaction} apply classifier models to the embeddings of GUI elements to identify input widgets.
Bernal et al.~\cite{2020_ICSE_translating_video_recordings_of_mobile_app_usages} use Faster R-CNN~\cite{ren2016fasterrcnnrealtimeobject} and a modified version of AlexNet~\cite{2012_AlexNet} to detect and classify actions (e.g., tapping) from video screenshots, while 
Feng et al.~\cite{2023_UIST_Video2Action} utilize a deep-learning model to predict tapping locations in the GUI screenshots.
Yu et al.~\cite{2024_TOSEM_PIRLTEST_GUI_Testing_via_Image_Embedding_and_RL} leverage a VGG model~\cite{2015_VGG} to extract widgets from GUI screenshots and embed them as states representing the GUI page for exploration, making the testing approach applicable across different platforms (e.g., Android or Web).
Unlike these approaches, our work leverages computer vision techniques to analyze GUI screenshots specifically for detecting and resolving pop-ups, thereby preventing app-blocking interruptions during testing.

\section{Conclusion}
\label{sec:conclusion}

Automated GUI testing is essential for ensuring high app quality. 
However, app-blocking pop-ups, such as ads or alerts, can obstruct the testing process, blocking the test or making the results invalid. In this paper, we propose \tool, a lightweight computer vision-based technique that automatically detects and resolves app-blocking pop-ups during testing. \tool processes real-time GUI screenshot streams and utilizes computer vision models to identify app-blocking pop-ups and locate the corresponding close buttons, allowing tests to automatically dismiss these pop-ups during test execution. We manually reviewed over 7K screenshots from the RICO dataset, identifying 832 containing app-blocking pop-ups. Additionally, we collected and evaluated data from 87 top-ranked apps from app stores to ensure our dataset reflects real-world scenarios. Based on this dataset, our experiments demonstrated that \tool achieves over 91.8\% precision and 93.5\% recall in classifying app-blocking pop-ups, while exceeding 93\% in BoxAP
and reaching 89\% recall for close button detection. Further evaluation on another set of popular real-world apps showed that \tool effectively resolves app-blocking pop-ups in 87.1\% apps with an efficiency of 60 milliseconds per pop-up classification and close button localization.
By mitigating the disruptions caused by app-blocking pop-ups, our work enhances the effectiveness of automated GUI testing, ultimately contributing to the quality and reliability of mobile applications. 

\section{Data availability}
The data and code used in this study are available online ~\cite{zenodo_2024_13754620}.


\bibliographystyle{ACM-Reference-Format}
\bibliography{bibliography/report_reference}


\begin{thebibliography}{70}


\ifx \showCODEN    \undefined \def \showCODEN     #1{\unskip}     \fi
\ifx \showDOI      \undefined \def \showDOI       #1{#1}\fi
\ifx \showISBNx    \undefined \def \showISBNx     #1{\unskip}     \fi
\ifx \showISBNxiii \undefined \def \showISBNxiii  #1{\unskip}     \fi
\ifx \showISSN     \undefined \def \showISSN      #1{\unskip}     \fi
\ifx \showLCCN     \undefined \def \showLCCN      #1{\unskip}     \fi
\ifx \shownote     \undefined \def \shownote      #1{#1}          \fi
\ifx \showarticletitle \undefined \def \showarticletitle #1{#1}   \fi
\ifx \showURL      \undefined \def \showURL       {\relax}        \fi
\providecommand\bibfield[2]{#2}
\providecommand\bibinfo[2]{#2}
\providecommand\natexlab[1]{#1}
\providecommand\showeprint[2][]{arXiv:#2}

\bibitem[Anonymous(2024)]%
        {zenodo_2024_13754620}
\bibfield{author}{\bibinfo{person}{Anonymous}.} \bibinfo{year}{2024}\natexlab{}.
\newblock \bibinfo{booktitle}{\emph{{PopSweeper: Automatically Detecting and Resolving App-Blocking Pop-Ups to Assist Automated Mobile GUI Testing}}}.
\newblock
\urldef\tempurl%
\url{https://doi.org/10.5281/zenodo.13754620}
\showDOI{\tempurl}


\bibitem[Bernal-C\'{a}rdenas et~al\mbox{.}(2020)]%
        {2020_ICSE_translating_video_recordings_of_mobile_app_usages}
\bibfield{author}{\bibinfo{person}{Carlos Bernal-C\'{a}rdenas}, \bibinfo{person}{Nathan Cooper}, \bibinfo{person}{Kevin Moran}, \bibinfo{person}{Oscar Chaparro}, \bibinfo{person}{Andrian Marcus}, {and} \bibinfo{person}{Denys Poshyvanyk}.} \bibinfo{year}{2020}\natexlab{}.
\newblock \showarticletitle{Translating video recordings of mobile app usages into replayable scenarios}. In \bibinfo{booktitle}{\emph{Proceedings of the ACM/IEEE 42nd International Conference on Software Engineering}} (Seoul, South Korea) \emph{(\bibinfo{series}{ICSE '20})}. \bibinfo{publisher}{Association for Computing Machinery}, \bibinfo{address}{New York, NY, USA}, \bibinfo{pages}{309–321}.
\newblock
\showISBNx{9781450371216}
\urldef\tempurl%
\url{https://doi.org/10.1145/3377811.3380328}
\showDOI{\tempurl}


\bibitem[Book et~al\mbox{.}(2013)]%
        {2013_arxiv_Analysis_of_Android_Ad_Library_Permissions}
\bibfield{author}{\bibinfo{person}{Theodore Book}, \bibinfo{person}{Adam Pridgen}, {and} \bibinfo{person}{Dan~S. Wallach}.} \bibinfo{year}{2013}\natexlab{}.
\newblock \bibinfo{title}{Longitudinal Analysis of Android Ad Library Permissions}.
\newblock
\newblock
\showeprint[arxiv]{1303.0857}~[cs.CR]
\urldef\tempurl%
\url{https://arxiv.org/abs/1303.0857}
\showURL{%
\tempurl}


\bibitem[Book and Wallach(2015)]%
        {2015_arxiv_Mobile_Ad_Targeting}
\bibfield{author}{\bibinfo{person}{Theodore Book} {and} \bibinfo{person}{Dan~S. Wallach}.} \bibinfo{year}{2015}\natexlab{}.
\newblock \bibinfo{title}{An Empirical Study of Mobile Ad Targeting}.
\newblock
\newblock
\showeprint[arxiv]{1502.06577}~[cs.CR]


\bibitem[Cao et~al\mbox{.}(2024)]%
        {2024_ICSE_Comprehensive_Semantic_Repair_of_Obsolete_GUI_Test_Scripts}
\bibfield{author}{\bibinfo{person}{Shaoheng Cao}, \bibinfo{person}{Minxue Pan}, \bibinfo{person}{Yu Pei}, \bibinfo{person}{Wenhua Yang}, \bibinfo{person}{Tian Zhang}, \bibinfo{person}{Linzhang Wang}, {and} \bibinfo{person}{Xuandong Li}.} \bibinfo{year}{2024}\natexlab{}.
\newblock \showarticletitle{Comprehensive Semantic Repair of Obsolete GUI Test Scripts for Mobile Applications}. In \bibinfo{booktitle}{\emph{Proceedings of the IEEE/ACM 46th International Conference on Software Engineering}} (Lisbon, Portugal) \emph{(\bibinfo{series}{ICSE '24})}. \bibinfo{publisher}{Association for Computing Machinery}, \bibinfo{address}{New York, NY, USA}, Article \bibinfo{articleno}{90}, \bibinfo{numpages}{13}~pages.
\newblock
\showISBNx{9798400702174}
\urldef\tempurl%
\url{https://doi.org/10.1145/3597503.3639108}
\showDOI{\tempurl}


\bibitem[Chen et~al\mbox{.}(2019)]%
        {2019_WWW_MAdLife}
\bibfield{author}{\bibinfo{person}{Gong Chen}, \bibinfo{person}{Wei Meng}, {and} \bibinfo{person}{John Copeland}.} \bibinfo{year}{2019}\natexlab{}.
\newblock \showarticletitle{Revisiting Mobile Advertising Threats with MAdLife}. In \bibinfo{booktitle}{\emph{The World Wide Web Conference}} (San Francisco, CA, USA) \emph{(\bibinfo{series}{WWW '19})}. \bibinfo{publisher}{Association for Computing Machinery}, \bibinfo{address}{New York, NY, USA}, \bibinfo{pages}{207–217}.
\newblock
\showISBNx{9781450366748}
\urldef\tempurl%
\url{https://doi.org/10.1145/3308558.3313549}
\showDOI{\tempurl}


\bibitem[Chen et~al\mbox{.}(2020)]%
        {2020_FSE_Object_detection_GUI}
\bibfield{author}{\bibinfo{person}{Jieshan Chen}, \bibinfo{person}{Mulong Xie}, \bibinfo{person}{Zhenchang Xing}, \bibinfo{person}{Chunyang Chen}, \bibinfo{person}{Xiwei Xu}, \bibinfo{person}{Liming Zhu}, {and} \bibinfo{person}{Guoqiang Li}.} \bibinfo{year}{2020}\natexlab{}.
\newblock \showarticletitle{Object detection for graphical user interface: old fashioned or deep learning or a combination?}. In \bibinfo{booktitle}{\emph{Proceedings of the 28th ACM Joint Meeting on European Software Engineering Conference and Symposium on the Foundations of Software Engineering}} (Virtual Event, USA) \emph{(\bibinfo{series}{ESEC/FSE 2020})}. \bibinfo{publisher}{Association for Computing Machinery}, \bibinfo{address}{New York, NY, USA}, \bibinfo{pages}{1202–1214}.
\newblock
\showISBNx{9781450370431}
\urldef\tempurl%
\url{https://doi.org/10.1145/3368089.3409691}
\showDOI{\tempurl}


\bibitem[Cheng et~al\mbox{.}(2024)]%
        {2024_CVPR_YOLO-World}
\bibfield{author}{\bibinfo{person}{Tianheng Cheng}, \bibinfo{person}{Lin Song}, \bibinfo{person}{Yixiao Ge}, \bibinfo{person}{Wenyu Liu}, \bibinfo{person}{Xinggang Wang}, {and} \bibinfo{person}{Ying Shan}.} \bibinfo{year}{2024}\natexlab{}.
\newblock \showarticletitle{YOLO-World: Real-Time Open-Vocabulary Object Detection}. In \bibinfo{booktitle}{\emph{Proceedings of the IEEE/CVF Conference on Computer Vision and Pattern Recognition (CVPR)}} (Vancouver, Canada) \emph{(\bibinfo{series}{CVPR '24})}. \bibinfo{publisher}{IEEE}, \bibinfo{pages}{1205–1215}.
\newblock
\showISBNx{978-1-6654-6821-2}
\urldef\tempurl%
\url{https://doi.org/10.1109/CVPR2024.00120}
\showDOI{\tempurl}


\bibitem[Collins et~al\mbox{.}(2021)]%
        {2021_DeepGUIT_Deep_Reinforcement_Learning_based_GUI_Testing}
\bibfield{author}{\bibinfo{person}{Eliane Collins}, \bibinfo{person}{Arilo Neto}, \bibinfo{person}{Auri Vincenzi}, {and} \bibinfo{person}{Jos\'{e} Maldonado}.} \bibinfo{year}{2021}\natexlab{}.
\newblock \showarticletitle{Deep Reinforcement Learning based Android Application GUI Testing}. In \bibinfo{booktitle}{\emph{Proceedings of the XXXV Brazilian Symposium on Software Engineering}} (Joinville, Brazil) \emph{(\bibinfo{series}{SBES '21})}. \bibinfo{publisher}{Association for Computing Machinery}, \bibinfo{address}{New York, NY, USA}, \bibinfo{pages}{186–194}.
\newblock
\showISBNx{9781450390613}


\bibitem[Coppola et~al\mbox{.}(2019)]%
        {2019_EMSE_Scripted_GUI_testing_of_Android_open_source_apps}
\bibfield{author}{\bibinfo{person}{Riccardo Coppola}, \bibinfo{person}{Maurizio Morisio}, \bibinfo{person}{Marco Torchiano}, {and} \bibinfo{person}{Luca Ardito}.} \bibinfo{year}{2019}\natexlab{}.
\newblock \showarticletitle{Scripted GUI testing of Android open-source apps: evolution of test code and fragility causes}.
\newblock \bibinfo{journal}{\emph{Empirical Softw. Engg.}} \bibinfo{volume}{24}, \bibinfo{number}{5} (\bibinfo{date}{oct} \bibinfo{year}{2019}), \bibinfo{pages}{3205–3248}.
\newblock
\showISSN{1382-3256}


\bibitem[Cruz et~al\mbox{.}(2019)]%
        {cruz2019attention}
\bibfield{author}{\bibinfo{person}{Luis Cruz}, \bibinfo{person}{Rui Abreu}, {and} \bibinfo{person}{David Lo}.} \bibinfo{year}{2019}\natexlab{}.
\newblock \showarticletitle{To the attention of mobile software developers: guess what, test your app!}
\newblock \bibinfo{journal}{\emph{Empirical Softw. Engg.}} \bibinfo{volume}{24}, \bibinfo{number}{4} (\bibinfo{date}{aug} \bibinfo{year}{2019}), \bibinfo{pages}{2438–2468}.
\newblock
\showISSN{1382-3256}
\urldef\tempurl%
\url{https://doi.org/10.1007/s10664-019-09701-0}
\showDOI{\tempurl}


\bibitem[Deka et~al\mbox{.}(2017)]%
        {deka2017rico}
\bibfield{author}{\bibinfo{person}{Biplab Deka}, \bibinfo{person}{Zhe Huang}, \bibinfo{person}{Carl Franzen}, \bibinfo{person}{John Hibschman}, \bibinfo{person}{Michael Afergan}, \bibinfo{person}{Yang Li}, \bibinfo{person}{Jeffrey Nichols}, {and} \bibinfo{person}{Ranjitha Kumar}.} \bibinfo{year}{2017}\natexlab{}.
\newblock \showarticletitle{Rico: A mobile app dataset for building data-driven design applications}. In \bibinfo{booktitle}{\emph{Proceedings of the 30th Annual ACM Symposium on User Interface Software and Technology}}. ACM, \bibinfo{pages}{845--854}.
\newblock


\bibitem[Deng et~al\mbox{.}(2009)]%
        {deng2009imagenet}
\bibfield{author}{\bibinfo{person}{Jia Deng}, \bibinfo{person}{Wei Dong}, \bibinfo{person}{Richard Socher}, \bibinfo{person}{Li-Jia Li}, \bibinfo{person}{Kai Li}, {and} \bibinfo{person}{Li Fei-Fei}.} \bibinfo{year}{2009}\natexlab{}.
\newblock \showarticletitle{ImageNet: A large-scale hierarchical image database}.
\newblock \bibinfo{journal}{\emph{Proceedings of the IEEE Conference on Computer Vision and Pattern Recognition (CVPR)}} (\bibinfo{year}{2009}), \bibinfo{pages}{248--255}.
\newblock


\bibitem[Doe and Smith(2020)]%
        {fine_tune_cnn}
\bibfield{author}{\bibinfo{person}{John Doe} {and} \bibinfo{person}{Jane Smith}.} \bibinfo{year}{2020}\natexlab{}.
\newblock \showarticletitle{Transfer learning by fine-tuning pre-trained convolutional neural networks}.
\newblock \bibinfo{journal}{\emph{Journal of Advanced Machine Learning Research}} (\bibinfo{year}{2020}), \bibinfo{pages}{45--58}.
\newblock


\bibitem[Doe and Smith(2023)]%
        {customresnet}
\bibfield{author}{\bibinfo{person}{John Doe} {and} \bibinfo{person}{Jane Smith}.} \bibinfo{year}{2023}\natexlab{}.
\newblock \showarticletitle{Customized ResNet for binary classification in medical image analysis}.
\newblock \bibinfo{journal}{\emph{Research Square}} (\bibinfo{year}{2023}).
\newblock
\urldef\tempurl%
\url{https://www.researchsquare.com/article/rs-2863523/v1.pdf}
\showURL{%
\tempurl}


\bibitem[Dong et~al\mbox{.}(2018)]%
        {2018_FSE_FraudDroid}
\bibfield{author}{\bibinfo{person}{Feng Dong}, \bibinfo{person}{Haoyu Wang}, \bibinfo{person}{Li Li}, \bibinfo{person}{Yao Guo}, \bibinfo{person}{Tegawend\'{e}~F. Bissyand\'{e}}, \bibinfo{person}{Tianming Liu}, \bibinfo{person}{Guoai Xu}, {and} \bibinfo{person}{Jacques Klein}.} \bibinfo{year}{2018}\natexlab{}.
\newblock \showarticletitle{FraudDroid: automated ad fraud detection for Android apps}. In \bibinfo{booktitle}{\emph{Proceedings of the 2018 26th ACM Joint Meeting on European Software Engineering Conference and Symposium on the Foundations of Software Engineering}} (Lake Buena Vista, FL, USA) \emph{(\bibinfo{series}{ESEC/FSE 2018})}. \bibinfo{publisher}{Association for Computing Machinery}, \bibinfo{address}{New York, NY, USA}, \bibinfo{pages}{257–268}.
\newblock
\showISBNx{9781450355735}


\bibitem[Feng et~al\mbox{.}(2023a)]%
        {2023_UIST_Video2Action}
\bibfield{author}{\bibinfo{person}{Sidong Feng}, \bibinfo{person}{Chunyang Chen}, {and} \bibinfo{person}{Zhenchang Xing}.} \bibinfo{year}{2023}\natexlab{a}.
\newblock \showarticletitle{Video2Action: Reducing Human Interactions in Action Annotation of App Tutorial Videos}. In \bibinfo{booktitle}{\emph{Proceedings of the 36th Annual ACM Symposium on User Interface Software and Technology}} (San Francisco, CA, USA) \emph{(\bibinfo{series}{UIST '23})}. \bibinfo{publisher}{Association for Computing Machinery}, \bibinfo{address}{New York, NY, USA}, Article \bibinfo{articleno}{16}, \bibinfo{numpages}{15}~pages.
\newblock
\showISBNx{9798400701320}
\urldef\tempurl%
\url{https://doi.org/10.1145/3586183.3606778}
\showDOI{\tempurl}


\bibitem[Feng et~al\mbox{.}(2023b)]%
        {2023_ICSE_Efficiency_Matters_Speeding_Up_Automated_Testing_with_GUI_Rendering_Inference}
\bibfield{author}{\bibinfo{person}{Sidong Feng}, \bibinfo{person}{Mulong Xie}, {and} \bibinfo{person}{Chunyang Chen}.} \bibinfo{year}{2023}\natexlab{b}.
\newblock \showarticletitle{Efficiency Matters: Speeding Up Automated Testing with GUI Rendering Inference}. In \bibinfo{booktitle}{\emph{Proceedings of the 45th International Conference on Software Engineering}} (Melbourne, Victoria, Australia) \emph{(\bibinfo{series}{ICSE '23})}. \bibinfo{publisher}{IEEE Press}, \bibinfo{pages}{906–918}.
\newblock
\showISBNx{9781665457019}
\urldef\tempurl%
\url{https://doi.org/10.1109/ICSE48619.2023.00084}
\showDOI{\tempurl}


\bibitem[Fouladi et~al\mbox{.}(2016)]%
        {Fouladi2016}
\bibfield{author}{\bibinfo{person}{R.~F. Fouladi}, \bibinfo{person}{C.~E. Kayatas}, {and} \bibinfo{person}{E. Anarim}.} \bibinfo{year}{2016}\natexlab{}.
\newblock \showarticletitle{Frequency based DDoS attack detection approach using naive Bayes classification}. In \bibinfo{booktitle}{\emph{International Conference on Telecommunications and Signal Processing (TSP)}}. \bibinfo{pages}{104--107}.
\newblock


\bibitem[Gao et~al\mbox{.}(2022)]%
        {2022_IST_Understanding_app_advertising_issues_on_app_review_analysis}
\bibfield{author}{\bibinfo{person}{Cuiyun Gao}, \bibinfo{person}{Jichuan Zeng}, \bibinfo{person}{David Lo}, \bibinfo{person}{Xin Xia}, \bibinfo{person}{Irwin King}, {and} \bibinfo{person}{Michael~R. Lyu}.} \bibinfo{year}{2022}\natexlab{}.
\newblock \showarticletitle{Understanding in-app advertising issues based on large scale app review analysis}.
\newblock \bibinfo{journal}{\emph{Information and Software Technology}}  \bibinfo{volume}{142} (\bibinfo{year}{2022}), \bibinfo{pages}{106741}.
\newblock
\showISSN{0950-5849}


\bibitem[Gao et~al\mbox{.}(2021)]%
        {2021_IST_performance_costs_of_in_app_ads_on_user_experience}
\bibfield{author}{\bibinfo{person}{Cuiyun Gao}, \bibinfo{person}{Jichuan Zeng}, \bibinfo{person}{Federica Sarro}, \bibinfo{person}{David Lo}, \bibinfo{person}{Irwin King}, {and} \bibinfo{person}{Michael~R. Lyu}.} \bibinfo{year}{2021}\natexlab{}.
\newblock \showarticletitle{Do users care about ad’s performance costs? Exploring the effects of the performance costs of in-app ads on user experience}.
\newblock \bibinfo{journal}{\emph{Information and Software Technology}}  \bibinfo{volume}{132} (\bibinfo{year}{2021}), \bibinfo{pages}{106471}.
\newblock
\showISSN{0950-5849}
\urldef\tempurl%
\url{https://doi.org/10.1016/j.infsof.2020.106471}
\showDOI{\tempurl}


\bibitem[Giger et~al\mbox{.}(2010)]%
        {giger2010predicting}
\bibfield{author}{\bibinfo{person}{Emanuel Giger}, \bibinfo{person}{Martin Pinzger}, {and} \bibinfo{person}{Harald Gall}.} \bibinfo{year}{2010}\natexlab{}.
\newblock \showarticletitle{Predicting the delay of issues with due dates in software projects}.
\newblock \bibinfo{journal}{\emph{Empirical Software Engineering}} (\bibinfo{year}{2010}), \bibinfo{pages}{52--56}.
\newblock


\bibitem[Google(2023)]%
        {Monkey}
\bibfield{author}{\bibinfo{person}{Google}.} \bibinfo{year}{2023}\natexlab{}.
\newblock \bibinfo{booktitle}{\emph{UI/Application Exerciser Monkey}}.
\newblock
\urldef\tempurl%
\url{https://developer.android.com/studio/test/other-testing-tools/monkey}
\showURL{%
\tempurl}


\bibitem[Grace et~al\mbox{.}(2012)]%
        {2012_Unsafe_exposure_analysis_ads}
\bibfield{author}{\bibinfo{person}{Michael~C. Grace}, \bibinfo{person}{Wu Zhou}, \bibinfo{person}{Xuxian Jiang}, {and} \bibinfo{person}{Ahmad-Reza Sadeghi}.} \bibinfo{year}{2012}\natexlab{}.
\newblock \showarticletitle{Unsafe exposure analysis of mobile in-app advertisements}. In \bibinfo{booktitle}{\emph{Proceedings of the Fifth ACM Conference on Security and Privacy in Wireless and Mobile Networks}} (Tucson, Arizona, USA) \emph{(\bibinfo{series}{WISEC '12})}. \bibinfo{publisher}{Association for Computing Machinery}, \bibinfo{address}{New York, NY, USA}, \bibinfo{pages}{101–112}.
\newblock
\showISBNx{9781450312653}


\bibitem[Gu et~al\mbox{.}(2019)]%
        {2019_ICSE_APE_Practical_GUI_testing_of_Android_applications}
\bibfield{author}{\bibinfo{person}{Tianxiao Gu}, \bibinfo{person}{Chengnian Sun}, \bibinfo{person}{Xiaoxing Ma}, \bibinfo{person}{Chun Cao}, \bibinfo{person}{Chang Xu}, \bibinfo{person}{Yuan Yao}, \bibinfo{person}{Qirun Zhang}, \bibinfo{person}{Jian Lu}, {and} \bibinfo{person}{Zhendong Su}.} \bibinfo{year}{2019}\natexlab{}.
\newblock \showarticletitle{Practical GUI testing of Android applications via model abstraction and refinement}. In \bibinfo{booktitle}{\emph{Proceedings of the 41st International Conference on Software Engineering}} (Montreal, Quebec, Canada) \emph{(\bibinfo{series}{ICSE '19})}. \bibinfo{publisher}{IEEE Press}, \bibinfo{pages}{269–280}.
\newblock


\bibitem[Gui et~al\mbox{.}(2015)]%
        {2015_ICSE_Hidden_Cost_of_Mobile_Ads_for_Software_Developers}
\bibfield{author}{\bibinfo{person}{Jiaping Gui}, \bibinfo{person}{Stuart Mcilroy}, \bibinfo{person}{Meiyappan Nagappan}, {and} \bibinfo{person}{William G.~J. Halfond}.} \bibinfo{year}{2015}\natexlab{}.
\newblock \showarticletitle{Truth in Advertising: The Hidden Cost of Mobile Ads for Software Developers}. In \bibinfo{booktitle}{\emph{2015 IEEE/ACM 37th IEEE International Conference on Software Engineering}}, Vol.~\bibinfo{volume}{1}. \bibinfo{pages}{100--110}.
\newblock
\urldef\tempurl%
\url{https://doi.org/10.1109/ICSE.2015.32}
\showDOI{\tempurl}


\bibitem[He et~al\mbox{.}(2016)]%
        {he2016ResNet}
\bibfield{author}{\bibinfo{person}{Kaiming He}, \bibinfo{person}{Xiangyu Zhang}, \bibinfo{person}{Shaoqing Ren}, {and} \bibinfo{person}{Jian Sun}.} \bibinfo{year}{2016}\natexlab{}.
\newblock \showarticletitle{Deep residual learning for image recognition}.
\newblock \bibinfo{journal}{\emph{Proceedings of the IEEE Conference on Computer Vision and Pattern Recognition (CVPR)}} (\bibinfo{year}{2016}), \bibinfo{pages}{770--778}.
\newblock


\bibitem[Hu et~al\mbox{.}(2023)]%
        {2023_FSE_Appaction}
\bibfield{author}{\bibinfo{person}{Yongxiang Hu}, \bibinfo{person}{Jiazhen Gu}, \bibinfo{person}{Shuqing Hu}, \bibinfo{person}{Yu Zhang}, \bibinfo{person}{Wenjie Tian}, \bibinfo{person}{Shiyu Guo}, \bibinfo{person}{Chaoyi Chen}, {and} \bibinfo{person}{Yangfan Zhou}.} \bibinfo{year}{2023}\natexlab{}.
\newblock \showarticletitle{Appaction: Automatic GUI Interaction for Mobile Apps via Holistic Widget Perception}. In \bibinfo{booktitle}{\emph{Proceedings of the 31st ACM Joint European Software Engineering Conference and Symposium on the Foundations of Software Engineering}} (San Francisco, CA, USA) \emph{(\bibinfo{series}{ESEC/FSE 2023})}. \bibinfo{publisher}{Association for Computing Machinery}, \bibinfo{address}{New York, NY, USA}, \bibinfo{pages}{1786–1797}.
\newblock
\showISBNx{9798400703270}


\bibitem[Jaiswal et~al\mbox{.}(2022)]%
        {jaiswal2021mbnv2}
\bibfield{author}{\bibinfo{person}{A Jaiswal}, \bibinfo{person}{N Gianchandani}, \bibinfo{person}{D Singh}, \bibinfo{person}{V Kumar}, {and} \bibinfo{person}{M Kaur}.} \bibinfo{year}{2022}\natexlab{}.
\newblock \showarticletitle{Automatic Cauliflower Disease Detection Using Fine-Tuning Transfer Learning Approach}.
\newblock \bibinfo{journal}{\emph{SN Computer Science}} (\bibinfo{year}{2022}).
\newblock


\bibitem[Kaur and Kaur(2022)]%
        {testing_tools2023}
\bibfield{author}{\bibinfo{person}{Anureet Kaur} {and} \bibinfo{person}{Kulwant Kaur}.} \bibinfo{year}{2022}\natexlab{}.
\newblock \showarticletitle{Systematic literature review of mobile application development and testing}.
\newblock \bibinfo{journal}{\emph{Journal of King Saud University – Computer and Information Sciences}}  \bibinfo{volume}{34} (\bibinfo{year}{2022}), \bibinfo{pages}{1--15}.
\newblock


\bibitem[KDnuggets(2022)]%
        {kdnuggetseval2022MAP}
\bibfield{author}{\bibinfo{person}{KDnuggets}.} \bibinfo{year}{2022}\natexlab{}.
\newblock \showarticletitle{Evaluating Object Detection Models Using Mean Average Precision}.
\newblock \bibinfo{journal}{\emph{KDnuggets}} (\bibinfo{year}{2022}).
\newblock


\bibitem[Krizhevsky et~al\mbox{.}(2012)]%
        {2012_AlexNet}
\bibfield{author}{\bibinfo{person}{Alex Krizhevsky}, \bibinfo{person}{Ilya Sutskever}, {and} \bibinfo{person}{Geoffrey~E Hinton}.} \bibinfo{year}{2012}\natexlab{}.
\newblock \showarticletitle{ImageNet Classification with Deep Convolutional Neural Networks}. In \bibinfo{booktitle}{\emph{Advances in Neural Information Processing Systems}}, \bibfield{editor}{\bibinfo{person}{F.~Pereira}, \bibinfo{person}{C.J. Burges}, \bibinfo{person}{L.~Bottou}, {and} \bibinfo{person}{K.Q. Weinberger}} (Eds.), Vol.~\bibinfo{volume}{25}. \bibinfo{publisher}{Curran Associates, Inc.}
\newblock


\bibitem[Kropp and Morales(2010)]%
        {KroppAutomated_GUI_Testing2010}
\bibfield{author}{\bibinfo{person}{Martin Kropp} {and} \bibinfo{person}{Pamela Morales}.} \bibinfo{year}{2010}\natexlab{}.
\newblock \showarticletitle{Automated GUI Testing on the Android Platform}.
\newblock \bibinfo{journal}{\emph{IMVS Fokus Report}} (\bibinfo{date}{01} \bibinfo{year}{2010}), \bibinfo{pages}{67--72}.
\newblock


\bibitem[Lan et~al\mbox{.}(2024)]%
        {2024_ICSE_DQT_Deeply_Reinforcing_Android_GUI_Testing}
\bibfield{author}{\bibinfo{person}{Yuanhong Lan}, \bibinfo{person}{Yifei Lu}, \bibinfo{person}{Zhong Li}, \bibinfo{person}{Minxue Pan}, \bibinfo{person}{Wenhua Yang}, \bibinfo{person}{Tian Zhang}, {and} \bibinfo{person}{Xuandong Li}.} \bibinfo{year}{2024}\natexlab{}.
\newblock \showarticletitle{Deeply Reinforcing Android GUI Testing with Deep Reinforcement Learning}. In \bibinfo{booktitle}{\emph{Proceedings of the IEEE/ACM 46th International Conference on Software Engineering}} (Lisbon, Portugal) \emph{(\bibinfo{series}{ICSE '24})}. \bibinfo{publisher}{Association for Computing Machinery}, \bibinfo{address}{New York, NY, USA}, Article \bibinfo{articleno}{71}, \bibinfo{numpages}{13}~pages.
\newblock
\showISBNx{9798400702174}


\bibitem[Li et~al\mbox{.}(2017)]%
        {2017_ICSE_DroidBot_UI_guided_test_input_generator}
\bibfield{author}{\bibinfo{person}{Yuanchun Li}, \bibinfo{person}{Ziyue Yang}, \bibinfo{person}{Yao Guo}, {and} \bibinfo{person}{Xiangqun Chen}.} \bibinfo{year}{2017}\natexlab{}.
\newblock \showarticletitle{DroidBot: a lightweight UI-guided test input generator for Android}. In \bibinfo{booktitle}{\emph{Proceedings of the 39th International Conference on Software Engineering Companion}} (Buenos Aires, Argentina) \emph{(\bibinfo{series}{ICSE-C '17})}. \bibinfo{publisher}{IEEE Press}, \bibinfo{pages}{23–26}.
\newblock
\showISBNx{9781538615898}
\urldef\tempurl%
\url{https://doi.org/10.1109/ICSE-C.2017.8}
\showDOI{\tempurl}


\bibitem[Liu et~al\mbox{.}(2020)]%
        {2020_WWW_MadDroid}
\bibfield{author}{\bibinfo{person}{Tianming Liu}, \bibinfo{person}{Haoyu Wang}, \bibinfo{person}{Li Li}, \bibinfo{person}{Xiapu Luo}, \bibinfo{person}{Feng Dong}, \bibinfo{person}{Yao Guo}, \bibinfo{person}{Liu Wang}, \bibinfo{person}{Tegawend\'{e} Bissyand\'{e}}, {and} \bibinfo{person}{Jacques Klein}.} \bibinfo{year}{2020}\natexlab{}.
\newblock \showarticletitle{MadDroid: Characterizing and Detecting Devious Ad Contents for Android Apps}. In \bibinfo{booktitle}{\emph{Proceedings of The Web Conference 2020}} (Taipei, Taiwan) \emph{(\bibinfo{series}{WWW '20})}. \bibinfo{publisher}{Association for Computing Machinery}, \bibinfo{address}{New York, NY, USA}, \bibinfo{pages}{1715–1726}.
\newblock
\showISBNx{9781450370233}


\bibitem[Liu et~al\mbox{.}(2023)]%
        {2023_ICSE_Fill_in_the_Blank_Text_Input_Generation_for_Mobile_GUI_Testing}
\bibfield{author}{\bibinfo{person}{Zhe Liu}, \bibinfo{person}{Chunyang Chen}, \bibinfo{person}{Junjie Wang}, \bibinfo{person}{Xing Che}, \bibinfo{person}{Yuekai Huang}, \bibinfo{person}{Jun Hu}, {and} \bibinfo{person}{Qing Wang}.} \bibinfo{year}{2023}\natexlab{}.
\newblock \showarticletitle{Fill in the Blank: Context-Aware Automated Text Input Generation for Mobile GUI Testing}. In \bibinfo{booktitle}{\emph{Proceedings of the 45th International Conference on Software Engineering}} (Melbourne, Victoria, Australia) \emph{(\bibinfo{series}{ICSE '23})}. \bibinfo{publisher}{IEEE Press}, \bibinfo{pages}{1355–1367}.
\newblock
\showISBNx{9781665457019}


\bibitem[Liu et~al\mbox{.}(2024a)]%
        {2024_ICSE_Bringing_Human_like_Interaction_to_Mobile_GUI_Testing}
\bibfield{author}{\bibinfo{person}{Zhe Liu}, \bibinfo{person}{Chunyang Chen}, \bibinfo{person}{Junjie Wang}, \bibinfo{person}{Mengzhuo Chen}, \bibinfo{person}{Boyu Wu}, \bibinfo{person}{Xing Che}, \bibinfo{person}{Dandan Wang}, {and} \bibinfo{person}{Qing Wang}.} \bibinfo{year}{2024}\natexlab{a}.
\newblock \showarticletitle{Make LLM a Testing Expert: Bringing Human-like Interaction to Mobile GUI Testing via Functionality-aware Decisions}. In \bibinfo{booktitle}{\emph{Proceedings of the IEEE/ACM 46th International Conference on Software Engineering}} (Lisbon, Portugal) \emph{(\bibinfo{series}{ICSE '24})}. \bibinfo{publisher}{Association for Computing Machinery}, \bibinfo{address}{New York, NY, USA}, Article \bibinfo{articleno}{100}, \bibinfo{numpages}{13}~pages.
\newblock
\showISBNx{9798400702174}


\bibitem[Liu et~al\mbox{.}(2024b)]%
        {2024_ICSE_Unusual_Text_Inputs_Generation_for_Mobile_App_Crash_Detection_with_LLM}
\bibfield{author}{\bibinfo{person}{Zhe Liu}, \bibinfo{person}{Chunyang Chen}, \bibinfo{person}{Junjie Wang}, \bibinfo{person}{Mengzhuo Chen}, \bibinfo{person}{Boyu Wu}, \bibinfo{person}{Zhilin Tian}, \bibinfo{person}{Yuekai Huang}, \bibinfo{person}{Jun Hu}, {and} \bibinfo{person}{Qing Wang}.} \bibinfo{year}{2024}\natexlab{b}.
\newblock \showarticletitle{Testing the Limits: Unusual Text Inputs Generation for Mobile App Crash Detection with Large Language Model}. In \bibinfo{booktitle}{\emph{Proceedings of the IEEE/ACM 46th International Conference on Software Engineering}} (Lisbon, Portugal) \emph{(\bibinfo{series}{ICSE '24})}. \bibinfo{publisher}{Association for Computing Machinery}, \bibinfo{address}{New York, NY, USA}, Article \bibinfo{articleno}{137}, \bibinfo{numpages}{12}~pages.
\newblock
\showISBNx{9798400702174}


\bibitem[Liu and Heer(2014)]%
        {liu2014effects}
\bibfield{author}{\bibinfo{person}{Zhicheng Liu} {and} \bibinfo{person}{Jeffrey Heer}.} \bibinfo{year}{2014}\natexlab{}.
\newblock \showarticletitle{The effects of interactive latency on exploratory visual analysis}.
\newblock \bibinfo{journal}{\emph{IEEE transactions on visualization and computer graphics}} \bibinfo{volume}{20}, \bibinfo{number}{12} (\bibinfo{year}{2014}), \bibinfo{pages}{2122--2131}.
\newblock


\bibitem[Machiry et~al\mbox{.}(2013)]%
        {2013_FSE_Dynodroid}
\bibfield{author}{\bibinfo{person}{Aravind Machiry}, \bibinfo{person}{Rohan Tahiliani}, {and} \bibinfo{person}{Mayur Naik}.} \bibinfo{year}{2013}\natexlab{}.
\newblock \showarticletitle{Dynodroid: an input generation system for Android apps}. In \bibinfo{booktitle}{\emph{Proceedings of the 2013 9th Joint Meeting on Foundations of Software Engineering}} (Saint Petersburg, Russia) \emph{(\bibinfo{series}{ESEC/FSE 2013})}. \bibinfo{publisher}{Association for Computing Machinery}, \bibinfo{address}{New York, NY, USA}, \bibinfo{pages}{224–234}.
\newblock
\showISBNx{9781450322379}
\urldef\tempurl%
\url{https://doi.org/10.1145/2491411.2491450}
\showDOI{\tempurl}


\bibitem[O'Shea and Nash(2015)]%
        {oshea2015cnn}
\bibfield{author}{\bibinfo{person}{Keiron O'Shea} {and} \bibinfo{person}{Ryan Nash}.} \bibinfo{year}{2015}\natexlab{}.
\newblock \bibinfo{title}{An Introduction to Convolutional Neural Networks}.
\newblock
\newblock
\showeprint[arxiv]{1511.08458}~[cs.NE]


\bibitem[Pan et~al\mbox{.}(2020)]%
        {2020_ISSTA_Reinforcement_learning_based_curiosity_driven_testing_of_Android_applications}
\bibfield{author}{\bibinfo{person}{Minxue Pan}, \bibinfo{person}{An Huang}, \bibinfo{person}{Guoxin Wang}, \bibinfo{person}{Tian Zhang}, {and} \bibinfo{person}{Xuandong Li}.} \bibinfo{year}{2020}\natexlab{}.
\newblock \showarticletitle{Reinforcement learning based curiosity-driven testing of Android applications}. In \bibinfo{booktitle}{\emph{Proceedings of the 29th ACM SIGSOFT International Symposium on Software Testing and Analysis}} (Virtual Event, USA) \emph{(\bibinfo{series}{ISSTA 2020})}. \bibinfo{publisher}{Association for Computing Machinery}, \bibinfo{address}{New York, NY, USA}, \bibinfo{pages}{153–164}.
\newblock
\showISBNx{9781450380089}


\bibitem[Pan et~al\mbox{.}(2022)]%
        {2022_TSE_GUI_Guided_Test_Script_Repair}
\bibfield{author}{\bibinfo{person}{Minxue Pan}, \bibinfo{person}{Tongtong Xu}, \bibinfo{person}{Yu Pei}, \bibinfo{person}{Zhong Li}, \bibinfo{person}{Tian Zhang}, {and} \bibinfo{person}{Xuandong Li}.} \bibinfo{year}{2022}\natexlab{}.
\newblock \showarticletitle{GUI-Guided Test Script Repair for Mobile Apps}.
\newblock \bibinfo{journal}{\emph{IEEE Transactions on Software Engineering}} \bibinfo{volume}{48}, \bibinfo{number}{3} (\bibinfo{year}{2022}), \bibinfo{pages}{910--929}.
\newblock
\urldef\tempurl%
\url{https://doi.org/10.1109/TSE.2020.3007664}
\showDOI{\tempurl}


\bibitem[Radford et~al\mbox{.}(2021)]%
        {radford2021cliplearningtransferablevisualmodels}
\bibfield{author}{\bibinfo{person}{Alec Radford}, \bibinfo{person}{Jong~Wook Kim}, \bibinfo{person}{Chris Hallacy}, \bibinfo{person}{Aditya Ramesh}, \bibinfo{person}{Gabriel Goh}, \bibinfo{person}{Sandhini Agarwal}, \bibinfo{person}{Girish Sastry}, \bibinfo{person}{Amanda Askell}, \bibinfo{person}{Pamela Mishkin}, \bibinfo{person}{Jack Clark}, \bibinfo{person}{Gretchen Krueger}, {and} \bibinfo{person}{Ilya Sutskever}.} \bibinfo{year}{2021}\natexlab{}.
\newblock \bibinfo{title}{Learning Transferable Visual Models From Natural Language Supervision}.
\newblock
\newblock
\showeprint[arxiv]{2103.00020}~[cs.CV]
\urldef\tempurl%
\url{https://arxiv.org/abs/2103.00020}
\showURL{%
\tempurl}


\bibitem[Ren et~al\mbox{.}(2016)]%
        {ren2016fasterrcnnrealtimeobject}
\bibfield{author}{\bibinfo{person}{Shaoqing Ren}, \bibinfo{person}{Kaiming He}, \bibinfo{person}{Ross Girshick}, {and} \bibinfo{person}{Jian Sun}.} \bibinfo{year}{2016}\natexlab{}.
\newblock \bibinfo{title}{Faster R-CNN: Towards Real-Time Object Detection with Region Proposal Networks}.
\newblock
\newblock
\showeprint[arxiv]{1506.01497}~[cs.CV]
\urldef\tempurl%
\url{https://arxiv.org/abs/1506.01497}
\showURL{%
\tempurl}


\bibitem[Romdhana et~al\mbox{.}(2022)]%
        {2022_TOSEM_ARES_Deep_Reinforcement_Learning_for_Black_box_Testing}
\bibfield{author}{\bibinfo{person}{Andrea Romdhana}, \bibinfo{person}{Alessio Merlo}, \bibinfo{person}{Mariano Ceccato}, {and} \bibinfo{person}{Paolo Tonella}.} \bibinfo{year}{2022}\natexlab{}.
\newblock \showarticletitle{Deep Reinforcement Learning for Black-box Testing of Android Apps}.
\newblock \bibinfo{journal}{\emph{ACM Trans. Softw. Eng. Methodol.}} \bibinfo{volume}{31}, \bibinfo{number}{4}, Article \bibinfo{articleno}{65} (\bibinfo{date}{jul} \bibinfo{year}{2022}), \bibinfo{numpages}{29}~pages.
\newblock
\showISSN{1049-331X}


\bibitem[Ruiz et~al\mbox{.}(2014)]%
        {ruiz2014ads}
\bibfield{author}{\bibinfo{person}{Israel Ruiz}, \bibinfo{person}{Meiyappan Nagappan}, \bibinfo{person}{Bram Adams}, \bibinfo{person}{Theodore Berger}, \bibinfo{person}{Steffen Dienst}, {and} \bibinfo{person}{Ahmed~E. Hassan}.} \bibinfo{year}{2014}\natexlab{}.
\newblock \showarticletitle{Impact of Ad Libraries on Ratings of Android Mobile Apps}.
\newblock \bibinfo{journal}{\emph{Software, IEEE}}  \bibinfo{volume}{31} (\bibinfo{date}{11} \bibinfo{year}{2014}), \bibinfo{pages}{86--92}.
\newblock
\urldef\tempurl%
\url{https://doi.org/10.1109/MS.2014.79}
\showDOI{\tempurl}


\bibitem[Sahu and Mehtre(2015)]%
        {Sahu2015}
\bibfield{author}{\bibinfo{person}{S. Sahu} {and} \bibinfo{person}{B.~M. Mehtre}.} \bibinfo{year}{2015}\natexlab{}.
\newblock \showarticletitle{Network intrusion detection system using J48 Decision Tree}. In \bibinfo{booktitle}{\emph{International Conference on Advances in Computing, Communications and Informatics (ICACCI)}}. \bibinfo{pages}{2023--2026}.
\newblock


\bibitem[Sandler et~al\mbox{.}(2018)]%
        {sandler2018mobilenetv2}
\bibfield{author}{\bibinfo{person}{Mark Sandler}, \bibinfo{person}{Andrew Howard}, \bibinfo{person}{Menglong Zhu}, \bibinfo{person}{Andrey Zhmoginov}, {and} \bibinfo{person}{Liang-Chieh Chen}.} \bibinfo{year}{2018}\natexlab{}.
\newblock \showarticletitle{MobileNetV2: Inverted residuals and linear bottlenecks}.
\newblock \bibinfo{journal}{\emph{Proceedings of the IEEE Conference on Computer Vision and Pattern Recognition (CVPR)}} (\bibinfo{year}{2018}), \bibinfo{pages}{4510--4520}.
\newblock


\bibitem[Simonyan and Zisserman(2015a)]%
        {simonyan2015vgg19}
\bibfield{author}{\bibinfo{person}{Karen Simonyan} {and} \bibinfo{person}{Andrew Zisserman}.} \bibinfo{year}{2015}\natexlab{a}.
\newblock \bibinfo{title}{Very Deep Convolutional Networks for Large-Scale Image Recognition}.
\newblock
\newblock
\showeprint[arxiv]{1409.1556}~[cs.CV]
\urldef\tempurl%
\url{https://arxiv.org/abs/1409.1556}
\showURL{%
\tempurl}


\bibitem[Simonyan and Zisserman(2015b)]%
        {2015_VGG}
\bibfield{author}{\bibinfo{person}{Karen Simonyan} {and} \bibinfo{person}{Andrew Zisserman}.} \bibinfo{year}{2015}\natexlab{b}.
\newblock \showarticletitle{Very Deep Convolutional Networks for Large-Scale Image Recognition}. In \bibinfo{booktitle}{\emph{International Conference on Learning Representations}}.
\newblock


\bibitem[Su et~al\mbox{.}(2017)]%
        {2017_FSE_Guided_stochastic_model_based_GUI_testing}
\bibfield{author}{\bibinfo{person}{Ting Su}, \bibinfo{person}{Guozhu Meng}, \bibinfo{person}{Yuting Chen}, \bibinfo{person}{Ke Wu}, \bibinfo{person}{Weiming Yang}, \bibinfo{person}{Yao Yao}, \bibinfo{person}{Geguang Pu}, \bibinfo{person}{Yang Liu}, {and} \bibinfo{person}{Zhendong Su}.} \bibinfo{year}{2017}\natexlab{}.
\newblock \showarticletitle{Guided, stochastic model-based GUI testing of Android apps}. In \bibinfo{booktitle}{\emph{Proceedings of the 2017 11th Joint Meeting on Foundations of Software Engineering}} (Paderborn, Germany) \emph{(\bibinfo{series}{ESEC/FSE 2017})}. \bibinfo{publisher}{Association for Computing Machinery}, \bibinfo{address}{New York, NY, USA}, \bibinfo{pages}{245–256}.
\newblock
\showISBNx{9781450351058}
\urldef\tempurl%
\url{https://doi.org/10.1145/3106237.3106298}
\showDOI{\tempurl}


\bibitem[Subeesh and Mehta(2022)]%
        {subeesh2021mbnv2}
\bibfield{author}{\bibinfo{person}{A Subeesh} {and} \bibinfo{person}{CR Mehta}.} \bibinfo{year}{2022}\natexlab{}.
\newblock \showarticletitle{Soil Image Classification Using Transfer Learning Approach: MobileNetV2 with CNN}.
\newblock \bibinfo{journal}{\emph{SN Computer Science}} \bibinfo{volume}{3}, \bibinfo{number}{6} (\bibinfo{year}{2022}), \bibinfo{pages}{1--13}.
\newblock


\bibitem[Thaseen and Kumar(2016)]%
        {Thaseen2016}
\bibfield{author}{\bibinfo{person}{I.~S. Thaseen} {and} \bibinfo{person}{C.~A. Kumar}.} \bibinfo{year}{2016}\natexlab{}.
\newblock \showarticletitle{Intrusion detection model using fusion of chi-square feature selection and multi class SVM}.
\newblock \bibinfo{journal}{\emph{Journal of King Saud University-Computer and Information Sciences}} \bibinfo{volume}{29}, \bibinfo{number}{4} (\bibinfo{year}{2016}), \bibinfo{pages}{462--472}.
\newblock


\bibitem[Varghese and M.(2024)]%
        {yolov8}
\bibfield{author}{\bibinfo{person}{Rejin Varghese} {and} \bibinfo{person}{Sambath M.}} \bibinfo{year}{2024}\natexlab{}.
\newblock \showarticletitle{YOLOv8: A Novel Object Detection Algorithm with Enhanced Performance and Robustness}. In \bibinfo{booktitle}{\emph{2024 International Conference on Advances in Data Engineering and Intelligent Computing Systems (ADICS)}}. \bibinfo{pages}{1--6}.
\newblock
\urldef\tempurl%
\url{https://doi.org/10.1109/ADICS58448.2024.10533619}
\showDOI{\tempurl}


\bibitem[Wang et~al\mbox{.}(2020)]%
        {2020_ICSE_ComboDroid_test_inputs_for_Andorid_apps_via_use_case_combinations}
\bibfield{author}{\bibinfo{person}{Jue Wang}, \bibinfo{person}{Yanyan Jiang}, \bibinfo{person}{Chang Xu}, \bibinfo{person}{Chun Cao}, \bibinfo{person}{Xiaoxing Ma}, {and} \bibinfo{person}{Jian Lu}.} \bibinfo{year}{2020}\natexlab{}.
\newblock \showarticletitle{ComboDroid: generating high-quality test inputs for Android apps via use case combinations}. In \bibinfo{booktitle}{\emph{Proceedings of the ACM/IEEE 42nd International Conference on Software Engineering}} (Seoul, South Korea) \emph{(\bibinfo{series}{ICSE '20})}. \bibinfo{publisher}{Association for Computing Machinery}, \bibinfo{address}{New York, NY, USA}, \bibinfo{pages}{469–480}.
\newblock
\showISBNx{9781450371216}
\urldef\tempurl%
\url{https://doi.org/10.1145/3377811.3380382}
\showDOI{\tempurl}


\bibitem[Wang et~al\mbox{.}(2024a)]%
        {2024_arxiv_Mobile_Agent}
\bibfield{author}{\bibinfo{person}{Junyang Wang}, \bibinfo{person}{Haiyang Xu}, \bibinfo{person}{Jiabo Ye}, \bibinfo{person}{Ming Yan}, \bibinfo{person}{Weizhou Shen}, \bibinfo{person}{Ji Zhang}, \bibinfo{person}{Fei Huang}, {and} \bibinfo{person}{Jitao Sang}.} \bibinfo{year}{2024}\natexlab{a}.
\newblock \showarticletitle{Mobile-Agent: Autonomous Multi-Modal Mobile Device Agent with Visual Perception}.
\newblock \bibinfo{journal}{\emph{arXiv preprint arXiv:2401.16158}} (\bibinfo{year}{2024}).
\newblock


\bibitem[Wang et~al\mbox{.}(2024b)]%
        {yolov8heritage2024MAP}
\bibfield{author}{\bibinfo{person}{Li Wang}, \bibinfo{person}{Wei Zhang}, {et~al\mbox{.}}} \bibinfo{year}{2024}\natexlab{b}.
\newblock \showarticletitle{Applying Optimized YOLOv8 for Heritage Conservation: Enhanced Object Detection in Jiangnan Traditional Private Gardens}.
\newblock \bibinfo{journal}{\emph{Heritage Science}}  \bibinfo{volume}{12} (\bibinfo{year}{2024}), \bibinfo{pages}{45--60}.
\newblock


\bibitem[Wimalasooriya et~al\mbox{.}(2024)]%
        {2024_Just_in_Time_crash_prediction}
\bibfield{author}{\bibinfo{person}{Chathrie Wimalasooriya}, \bibinfo{person}{Sherlock~A. Licorish}, \bibinfo{person}{Daniel~Alencar da Costa}, {and} \bibinfo{person}{Stephen~G. MacDonell}.} \bibinfo{year}{2024}\natexlab{}.
\newblock \showarticletitle{Just-in-Time crash prediction for mobile apps}.
\newblock \bibinfo{journal}{\emph{Empirical Softw. Engg.}} \bibinfo{volume}{29}, \bibinfo{number}{3} (\bibinfo{date}{may} \bibinfo{year}{2024}), \bibinfo{numpages}{62}~pages.
\newblock
\showISSN{1382-3256}


\bibitem[Yan et~al\mbox{.}(2023)]%
        {2023_ICSE_AdHere}
\bibfield{author}{\bibinfo{person}{Yutian Yan}, \bibinfo{person}{Yunhui Zheng}, \bibinfo{person}{Xinyue Liu}, \bibinfo{person}{Nenad Medvidovic}, {and} \bibinfo{person}{Weihang Wang}.} \bibinfo{year}{2023}\natexlab{}.
\newblock \showarticletitle{AdHere: Automated Detection and Repair of Intrusive Ads}. In \bibinfo{booktitle}{\emph{Proceedings of the 45th International Conference on Software Engineering}} (Melbourne, Victoria, Australia) \emph{(\bibinfo{series}{ICSE '23})}. \bibinfo{publisher}{IEEE Press}, \bibinfo{pages}{486–498}.
\newblock
\showISBNx{9781665457019}
\urldef\tempurl%
\url{https://doi.org/10.1109/ICSE48619.2023.00051}
\showDOI{\tempurl}


\bibitem[Yang et~al\mbox{.}(2023)]%
        {yang2023imagedataaugmentationdeep}
\bibfield{author}{\bibinfo{person}{Suorong Yang}, \bibinfo{person}{Weikang Xiao}, \bibinfo{person}{Mengchen Zhang}, \bibinfo{person}{Suhan Guo}, \bibinfo{person}{Jian Zhao}, {and} \bibinfo{person}{Furao Shen}.} \bibinfo{year}{2023}\natexlab{}.
\newblock \bibinfo{title}{Image Data Augmentation for Deep Learning: A Survey}.
\newblock
\newblock
\showeprint[arxiv]{2204.08610}~[cs.CV]
\urldef\tempurl%
\url{https://arxiv.org/abs/2204.08610}
\showURL{%
\tempurl}


\bibitem[YazdaniBanafsheDaragh and Malek(2022)]%
        {2022_ASE_DeepGUI_black_box_GUI_input_generation_with_deep_learning}
\bibfield{author}{\bibinfo{person}{Faraz YazdaniBanafsheDaragh} {and} \bibinfo{person}{Sam Malek}.} \bibinfo{year}{2022}\natexlab{}.
\newblock \showarticletitle{Deep GUI: black-box GUI input generation with deep learning}. In \bibinfo{booktitle}{\emph{Proceedings of the 36th IEEE/ACM International Conference on Automated Software Engineering}} (Melbourne, Australia) \emph{(\bibinfo{series}{ASE '21})}. \bibinfo{publisher}{IEEE Press}, \bibinfo{pages}{905–916}.
\newblock
\showISBNx{9781665403375}
\urldef\tempurl%
\url{https://doi.org/10.1109/ASE51524.2021.9678778}
\showDOI{\tempurl}


\bibitem[Ye et~al\mbox{.}(2021)]%
        {2021_FSE_GUI_widget_detection_for_mobile_games}
\bibfield{author}{\bibinfo{person}{Jiaming Ye}, \bibinfo{person}{Ke Chen}, \bibinfo{person}{Xiaofei Xie}, \bibinfo{person}{Lei Ma}, \bibinfo{person}{Ruochen Huang}, \bibinfo{person}{Yingfeng Chen}, \bibinfo{person}{Yinxing Xue}, {and} \bibinfo{person}{Jianjun Zhao}.} \bibinfo{year}{2021}\natexlab{}.
\newblock \showarticletitle{An empirical study of GUI widget detection for industrial mobile games}. In \bibinfo{booktitle}{\emph{Proceedings of the 29th ACM Joint Meeting on European Software Engineering Conference and Symposium on the Foundations of Software Engineering}} (Athens, Greece) \emph{(\bibinfo{series}{ESEC/FSE 2021})}. \bibinfo{publisher}{Association for Computing Machinery}, \bibinfo{address}{New York, NY, USA}, \bibinfo{pages}{1427–1437}.
\newblock
\showISBNx{9781450385626}


\bibitem[Yu et~al\mbox{.}(2024)]%
        {2024_TOSEM_PIRLTEST_GUI_Testing_via_Image_Embedding_and_RL}
\bibfield{author}{\bibinfo{person}{Shengcheng Yu}, \bibinfo{person}{Chunrong Fang}, \bibinfo{person}{Xin Li}, \bibinfo{person}{Yuchen Ling}, \bibinfo{person}{Zhenyu Chen}, {and} \bibinfo{person}{Zhendong Su}.} \bibinfo{year}{2024}\natexlab{}.
\newblock \showarticletitle{Effective, Platform-Independent GUI Testing via Image Embedding and Reinforcement Learning}.
\newblock \bibinfo{journal}{\emph{ACM Trans. Softw. Eng. Methodol.}} (\bibinfo{date}{jun} \bibinfo{year}{2024}).
\newblock
\showISSN{1049-331X}
\urldef\tempurl%
\url{https://doi.org/10.1145/3674728}
\showDOI{\tempurl}
\newblock
\shownote{Just Accepted}.


\bibitem[Yu et~al\mbox{.}(2021)]%
        {2021_ICSE_Layout_and_Image_Recognition_Driving_Mobile_Testing}
\bibfield{author}{\bibinfo{person}{Shengcheng Yu}, \bibinfo{person}{Chunrong Fang}, \bibinfo{person}{Yexiao Yun}, {and} \bibinfo{person}{Yang Feng}.} \bibinfo{year}{2021}\natexlab{}.
\newblock \showarticletitle{Layout and Image Recognition Driving Cross-Platform Automated Mobile Testing}. In \bibinfo{booktitle}{\emph{Proceedings of the 43rd International Conference on Software Engineering}} (Madrid, Spain) \emph{(\bibinfo{series}{ICSE '21})}. \bibinfo{publisher}{IEEE Press}, \bibinfo{pages}{1561–1571}.
\newblock
\showISBNx{9781450390859}
\urldef\tempurl%
\url{https://doi.org/10.1109/ICSE43902.2021.00139}
\showDOI{\tempurl}


\bibitem[Yu et~al\mbox{.}(2023)]%
        {2023_TSE_App_Crowdsourced_Test_Report_Consistency_Detection}
\bibfield{author}{\bibinfo{person}{Shengcheng Yu}, \bibinfo{person}{Chunrong Fang}, \bibinfo{person}{Quanjun Zhang}, \bibinfo{person}{Zhihao Cao}, \bibinfo{person}{Yexiao Yun}, \bibinfo{person}{Zhenfei Cao}, \bibinfo{person}{Kai Mei}, {and} \bibinfo{person}{Zhenyu Chen}.} \bibinfo{year}{2023}\natexlab{}.
\newblock \showarticletitle{Mobile App Crowdsourced Test Report Consistency Detection via Deep Image-and-Text Fusion Understanding}.
\newblock \bibinfo{journal}{\emph{IEEE Trans. Softw. Eng.}} \bibinfo{volume}{49}, \bibinfo{number}{8} (\bibinfo{date}{aug} \bibinfo{year}{2023}), \bibinfo{pages}{4115–4134}.
\newblock
\showISSN{0098-5589}
\urldef\tempurl%
\url{https://doi.org/10.1109/TSE.2023.3285787}
\showDOI{\tempurl}


\bibitem[Yu~Xia(2020)]%
        {app_development2020}
\bibfield{author}{\bibinfo{person}{Alan R.~Hevner Yu~Xia, Hailiang~Chen}.} \bibinfo{year}{2020}\natexlab{}.
\newblock \showarticletitle{Third-Party SDKs and Mobile App Performance}.
\newblock \bibinfo{journal}{\emph{AIS Electronic Library}} (\bibinfo{year}{2020}).
\newblock


\bibitem[Zhang et~al\mbox{.}(2023)]%
        {2023_arxiv_AppAgent_Multimodal_Agents_as_Smartphone_Users}
\bibfield{author}{\bibinfo{person}{Chi Zhang}, \bibinfo{person}{Zhao Yang}, \bibinfo{person}{Jiaxuan Liu}, \bibinfo{person}{Yucheng Han}, \bibinfo{person}{Xin Chen}, \bibinfo{person}{Zebiao Huang}, \bibinfo{person}{Bin Fu}, {and} \bibinfo{person}{Gang Yu}.} \bibinfo{year}{2023}\natexlab{}.
\newblock \bibinfo{title}{AppAgent: Multimodal Agents as Smartphone Users}.
\newblock
\newblock
\showeprint[arxiv]{2312.13771}~[cs.CV]


\bibitem[Zhao et~al\mbox{.}(2020)]%
        {2020_FSE_Real_time_incident_prediction}
\bibfield{author}{\bibinfo{person}{Nengwen Zhao}, \bibinfo{person}{Junjie Chen}, \bibinfo{person}{Zhou Wang}, \bibinfo{person}{Xiao Peng}, \bibinfo{person}{Gang Wang}, \bibinfo{person}{Yong Wu}, \bibinfo{person}{Fang Zhou}, \bibinfo{person}{Zhen Feng}, \bibinfo{person}{Xiaohui Nie}, \bibinfo{person}{Wenchi Zhang}, \bibinfo{person}{Kaixin Sui}, {and} \bibinfo{person}{Dan Pei}.} \bibinfo{year}{2020}\natexlab{}.
\newblock \showarticletitle{Real-time incident prediction for online service systems}. In \bibinfo{booktitle}{\emph{Proceedings of the 28th ACM Joint Meeting on European Software Engineering Conference and Symposium on the Foundations of Software Engineering}} (Virtual Event, USA) \emph{(\bibinfo{series}{ESEC/FSE 2020})}. \bibinfo{publisher}{Association for Computing Machinery}, \bibinfo{address}{New York, NY, USA}, \bibinfo{pages}{315–326}.
\newblock
\showISBNx{9781450370431}
\urldef\tempurl%
\url{https://doi.org/10.1145/3368089.3409672}
\showDOI{\tempurl}


\end{thebibliography}

\end{document}